\documentclass{aa}  
\usepackage{graphics}
\usepackage{graphicx}
\usepackage{multicol}
\usepackage[varg]{txfonts}

\usepackage{natbib,twoopt}
\usepackage{amsmath} 
\usepackage[breaklinks=true]{hyperref} 
\bibpunct{(}{)}{;}{a}{}{,} 
\makeatletter
\newcommandtwoopt{\citeads}[3][][]{\href{http://adsabs.harvard.edu/abs/#3}%
{\def\hyper@linkstart##1##2{}%
\let\hyper@linkend\@empty\citealp[#1][#2]{#3}}}
\newcommandtwoopt{\citepads}[3][][]{\href{http://adsabs.harvard.edu/abs/#3}%
{\def\hyper@linkstart##1##2{}%
\let\hyper@linkend\@empty\citep[#1][#2]{#3}}}
\newcommandtwoopt{\citetads}[3][][]{\href{http://adsabs.harvard.edu/abs/#3}%
{\def\hyper@linkstart##1##2{}%
\let\hyper@linkend\@empty\citet[#1][#2]{#3}}}
\newcommandtwoopt{\citeyearads}[3][][]%
{\href{http://adsabs.harvard.edu/abs/#3}
{\def\hyper@linkstart##1##2{}%
\let\hyper@linkend\@empty\citeyear[#1][#2]{#3}}}
\makeatother

\usepackage{color}
\definecolor{mygreen}{RGB}{0,128,0}

\hypersetup{colorlinks=true,linkcolor=blue,citecolor=blue,urlcolor=blue}

\def\barypmRA{-3619.9} 
\def\barypmdec{+693.8} 
\def\barypmerr{3.9} 
\def\baryvalpha{-22.9666} 
\def\baryvalphaerr{0.0193} 
\def\baryvdelta{+4.4019} 
\def\baryvdeltaerr{0.0037} 
\def\residualshiftA{0.025}
\def\residualshiftB{0.029}
\def\minSone{0.205} 
\def\errminSone{0.177} 
\def\dateSone{2021-05-03}
\def\PMRASone{-5} 
\def\PMdecSone{-12} 
\def\PMerrSone{12} 
\def\minStwo{1.643} 
\def\errminStwo{0.112} 
\def\dateStwo{2023-04-27}
\def\PMRAStwo{+2} 
\def\PMdecStwo{-2} 
\def\PMerrStwo{5} 
\def\minSthree{1.354} 
\def\errminSthree{0.186} 
\def\dateSthree{2023-12-12}
\def\PMRASthree{+1} 
\def\PMdecSthree{-4} 
\def\PMerrSthree{11} 
\def\minSfour{2.433} 
\def\errminSfour{0.119} 
\def\dateSfour{2024-10-26}
\def\PMRASfour{+6} 
\def\PMdecSfour{+8} 
\def\PMerrSfour{5} 
\def\minSfive{0.015} 
\def\errminSfive{0.135} 
\def\dateSfive{2028-05-06}
\def\PMRASfive{-5} 
\def\PMdecSfive{-5} 
\def\PMerrSfive{5} 
\def\minSsix{0.269} 
\def\errminSsix{0.277} 
\def\dateSsix{2031-05-25}
\def\PMRASsix{-6} 
\def\PMdecSsix{-1} 
\def\PMerrSsix{11} 

\begin{document} 
\title{Close stellar conjunctions of $\alpha$~Centauri A and B until 2050\thanks{Based on observations collected at the European Organisation for Astronomical Research in the Southern Hemisphere 
under ESO programs 272.C-5010, 077.C-0587, 080.C-0775(A), 081.C-0940(A), 082.C-0615(A) and 296.D-5032.}
}
\subtitle{An $m_K = 7.8$ star may enter the Einstein ring of $\alpha$\,Cen~A in 2028}
\titlerunning{Close stellar conjunctions with $\alpha$\,Centauri}
\authorrunning{P. Kervella et al.}

\author{
P.~Kervella\inst{1,2}
\and
F.~Mignard\inst{3}
\and
A.~M\'erand\inst{4}
\and
F.~Th\'evenin\inst{3}
}
\institute{
Unidad Mixta Internacional Franco-Chilena de Astronom\'{i}a (CNRS UMI 3386), Departamento de Astronom\'{i}a, Universidad de Chile, Camino El Observatorio 1515, Las Condes, Santiago, Chile, \email{pkervell@das.uchile.cl}.
\and
LESIA (UMR 8109), Observatoire de Paris, PSL Research University, CNRS, UPMC, Univ. Paris-Diderot, 5 Place Jules Janssen, 92195 Meudon, France, \email{pierre.kervella@obspm.fr}.
\and
Laboratoire Lagrange, Universit{\'e} Nice Sophia-Antipolis, Observatoire de la C{\^o}te d’Azur, CNRS, CS 34229, F-06304 Nice Cedex, France 
\and
European Southern Observatory, Alonso de C\'ordova 3107, Casilla 19001, Santiago 19, Chile.
}
\date{Received ; Accepted}

 
  \abstract
{
The rapid proper motion of the $\alpha$\,Cen pair ($\approx 3.7$\,arcsec\ yr$^{-1}$) and its location close to the galactic plane on a rich stellar background combine constructively to make them excellent candidates for close stellar conjunctions with more distant stars.
Adding new differential astrometry to archival data, we have refined the orbital parameters, barycentric proper motion and parallax of $\alpha$\,Cen and compute its apparent trajectory on sky over the coming decades.
Based on NTT/SUSI2, NTT/SOFI and VLT/NACO maps of the field stars around the trajectories of $\alpha$\,Cen A and B, we present a catalog of the expected close conjunctions until 2050.
An exceptional event will take place in early May 2028, when $\alpha$\,Cen~A will come within $\rho_\mathrm{min} = \minSfive \pm \errminSfive$\,arcseconds of the $m_K = 7.8$ star 2MASS 14392160-6049528 (hereafter S5).
In terms of impact parameter and contrast, this is the most favorable stellar conjunction of $\alpha$\,Cen within at least the next three decades.
With an angular diameter of $\theta_\mathrm{LD} = 0.47 \pm 0.05$\,mas, it is likely that S5 is a red giant or supergiant located at several kiloparsecs.
The approached stars will act as moving light probes in transmission through the environment of $\alpha$\,Cen.
The observation of these close conjunctions holds great promises to search for planets and other low mass objects in the $\alpha$\,Cen system using photometry and astrometry.
The relativistic displacement of the approached star images will be observable, with significant deflection angles in the milliarcsecond range.
The small impact parameter of the conjunction with S5 means that this star has a probability of $45\%$ of entering the Einstein ring of $\alpha$\,Cen~A.
The gravitational amplification of the flux of S5 could reach a factor five for the combination of the two lensed images.
The proper motion, orbital parameters and parallax of $\alpha$\,Cen will be measurable with an extreme accuracy from differential astrometry with the S stars.
This will be valuable, for example to prepare the recently announced \emph{Breakthrough Starshot} initiative to send interstellar nanocrafts to $\alpha$\,Centauri.
}


   \maketitle
%

\section{Introduction} 
 
The triple star $\alpha$ Centauri (\object{WDS J14396-6050AB}, \object{GJ559AB}) is our closest stellar neighbor. The main stars $\alpha$\,Cen A and B are located at a distance of only $d = 1.338 \pm 0.001$\,pc ($\pi = 747.17 \pm 0.61$\,mas; Sect.~\ref{orbit}).
Their orbital period is close to 80\,years, with a projected semi-major axis of $17.6\arcsec$.
The main components are dwarf stars of G2V (component A, \object{HD 128620}) and K1V (component B, \object{HD 128621}) spectral types, while the third member is the red dwarf \object{Proxima} (M5.5V, \object{GJ551}).
\object{Proxima} is actually slightly closer to Earth by approximately 7\,500\,AU, at $d = 1.302 \pm 0.002$\,pc ($\pi = 768.13 \pm 1.04$\,mas; \citeads{2014AJ....148...91L}).
The proximity of $\alpha$\,Cen A and B and their similarity to the Sun in terms of mass and spectral type make these stars extremely attractive from the standpoint of both stellar physics \citepads{2016MNRAS.460.1254B} and extrasolar planet research \citepads{2015MNRAS.450.2043D}.

As predicted by \citetads{1996AcA....46..291P} and discussed in details by \citetads{2008ApJ...684...59D, 2008ApJ...684...46D} the close conjunction of a foreground mass as $\alpha$\,Cen with a distant background star will create a gravitational mesolensing event.
Lists of stellar conjunctions have been predicted for example, by \citetads{2000ApJ...539..241S}, \citetads{2011A&A...536A..50P}, and for specific objects by \citetads{2012ApJ...749L...6L} (\object{VB 10}) and \citetads{2014ApJ...782...89S} (\object{Proxima}).
But to our knowledge, none of the predicted lensing events has been observed successfully due to the very small observational signature of the expected gravitational deflection.
Thanks to the large angular size of their Einstein rings and their fast displacement on a rich stellar background, $\alpha$\,Cen A and B are prime candidates for such events.

In this paper, we use new and archival observations of $\alpha$\,Cen to determine the conjunctions with background stars that will occur in the coming decades.
These events are relatively rare and usually difficult to predict as star catalogs are poorly populated in the immediate vicinity of such very bright stars.
Even in the deepest catalogs, only a handful of stars are typically present within $1\arcmin$ of $\alpha$\,Cen, and many are artefacts.
Measuring accurate astrometric and photometric parameters is also made difficult by the residual speckles and the spatially variable background.
More than a decade ago, two of us obtained deep observations of the field surrounding $\alpha$\,Cen using the (now decommissioned) NTT/SUSI2 instrument \citepads{2007A&A...464..373K}.
This allowed us to assemble a catalog of the field stars present within a few arcminutes of $\alpha$\,Cen.
While this catalog was far from complete within $\approx 30\arcsec$ of the two stars at the time, the very fast proper motion of $\alpha$\,Cen has now brought the pair inside a much better known star field.
As $\alpha$\,Cen is located almost exactly in the plane of the Milky Way, background stars are extremely numerous, particularly at infrared wavelengths where the interstellar extinction is weaker.
This has already been noted by \citetads{2006A&A...459..669K} who observed hundreds of stars within $\approx 20\arcsec$ of $\alpha$\,Cen using the VLT/NACO instrument.

We present in Sect.~\ref{observations} the new VLT/NACO observations that we obtained in March 2016 as well as archival data from VLT/NACO, NTT/SOFI and ALMA.
We take particular care to reference precisely the positions of $\alpha$\,Cen and the celestial coordinate frame of our catalog.
The trajectory of the $\alpha$\,Cen pair is derived in Sect.~\ref{astrometryAB}, in which we have also refined the orbital parameters, parallax and masses of the two components.
From our catalog of background stars, and taking into account the proper motion, the orbital motion and the parallactic motion of $\alpha$\,Cen, we have derived in Sect.~\ref{conjunctions} the list of stellar conjunctions that will occur with $\alpha$\,Cen A and B until 2050.
We identify six conjunctions up to 2031 that we discuss individually, focusing in particular on the approach of star S5 in 2028.
We estimate the angular diameters of the corresponding background sources, and obtain upper limits on theirs proper motions.
We discuss in Sect.~\ref{potential} the potential of the conjunctions in terms of gravitational deflection and amplification of the light from the approached background stars.
We also address the scientific potential of such observations to search for low mass objects in the $\alpha$\,Cen system.

\section{Observations and data reduction\label{observations}}

\subsection{NTT/SUSI2 and NTT/SOFI imaging}

\begin{figure*}[]
        \centering
        \includegraphics[width=\hsize]{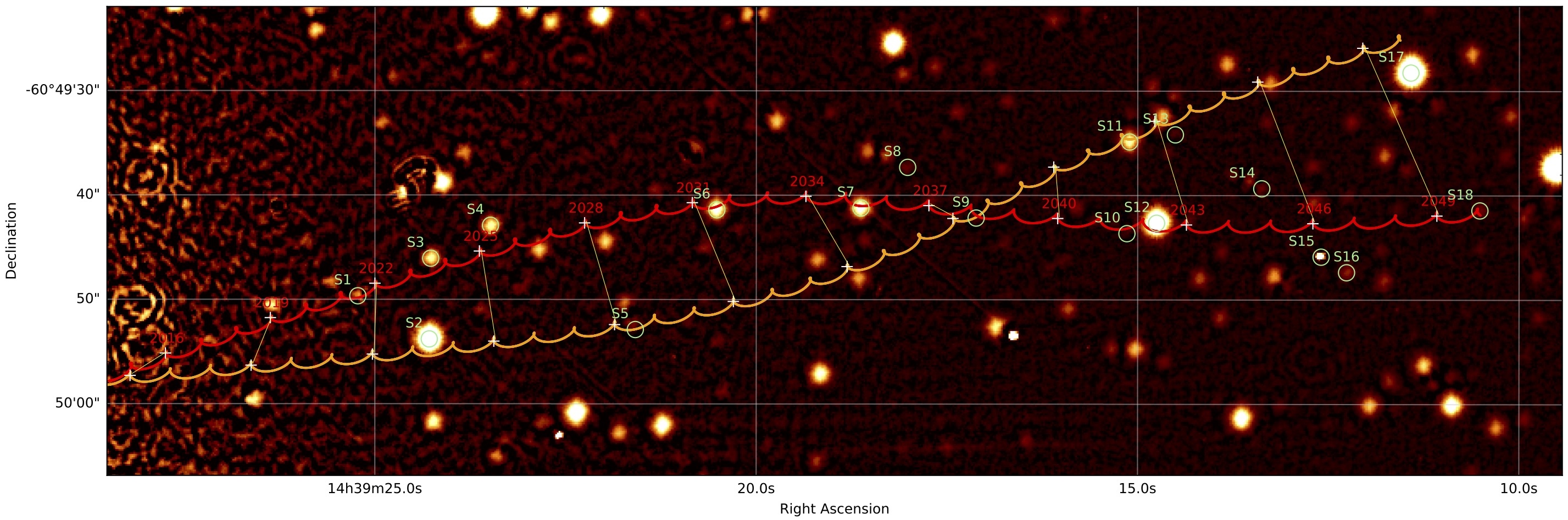}
        \caption{SUSI2 field in the $I$ band on 2 April 2004 and trajectories of $\alpha$\,Cen A (orange curve) and B (red curve) until 2050.
        The positions of the approached S stars are circled in green (Sect.~\ref{conjunction-list}).
        \label{susi2-field}}
\end{figure*}

The SUSI2 exposures of the field around $\alpha$\,Cen in the $VRIZ$ bands were presented by \citetads{2007A&A...464..373K}, where a catalog of the sources is also provided. This catalog is available through ViZieR with revised coordinates (see Sect.~\ref{referencing}) and improved photometry. An example image in the $I$ band is presented in Fig.~\ref{susi2-field}.
To complement these images, we retrieved from the ESO archive a series of exposures of $\alpha$\,Cen that were obtained using the NTT/SOFI instrument \citepads{1998Msngr..91....9M}.
These exposures were taken in March 2008, August 2008 and February 2009 in the infrared $J_s$ and $K_s$ bands for programs 080.C-0775(A), 081.C-0940(A) and 082.C-0615(A).
The exposures in the $K_s$ band are generally saturated on S2 and S5, and we therefore selected the best exposure in the $J_s$ band for which S2 and S5 are unsaturated. This image was obtained on 25 February 2009 at UT 08:32:38. The exposure time is $20 \times 1.2$\,s in the $J_s$ filter\footnote{\url{https://www.eso.org/sci/facilities/lasilla/instruments/sofi/inst/Imaging.html}} ($\lambda_0 = 1.24\,\mu$m, $\Delta \lambda = 0.16\,\mu$m) using the large field of view ($4.9\arcmin \times 4.9\arcmin$, $0.288\arcsec$\,pix$^{-1}$). The observatory seeing was $0.82\arcsec$ in the visible, and the observation took place at an airmass of 1.18. The raw image was classically corrected for flat field.

\subsection{VLT/NACO adaptive optics\label{naco-obs}}

\begin{figure}[ht]
        \centering
        \includegraphics[width=\hsize]{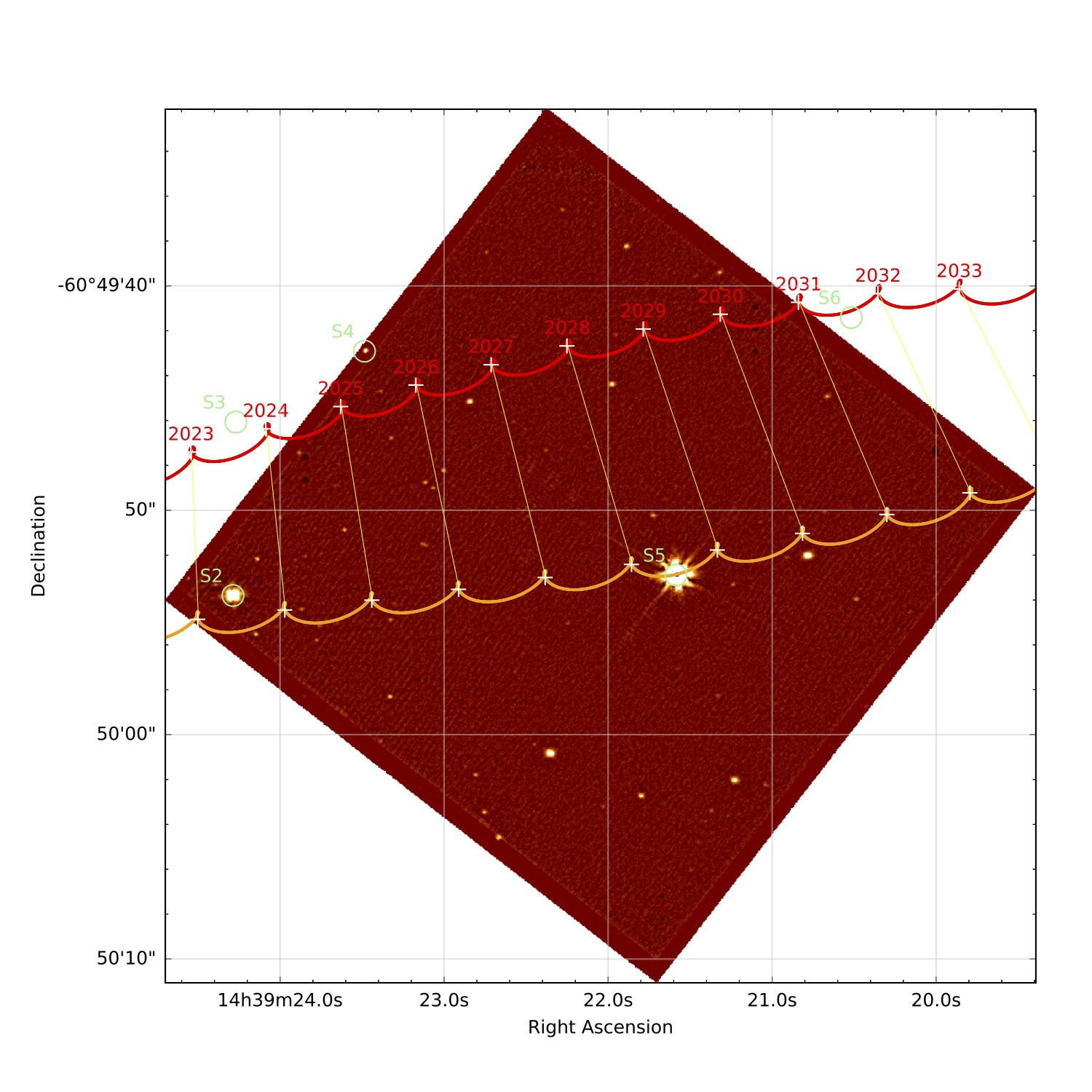}
        \caption{NACO exposure obtained on 27 March 2016 of the field that will be crossed by $\alpha$\,Cen between 2023 and 2032.\label{alfcen-naco}}
\end{figure}

We observed a field located in front of $\alpha$\,Cen in the direction of its proper motion on 27 March 2016 using the NACO instrument \citepads{2010SPIE.7736E..2NG} installed at Unit Telescope 1 of the Very Large Telescope (VLT). This instrument comprises the Nasmyth Adaptive Optics System \citepads[NAOS; ]{2003SPIE.4839..140R} and the CONICA infrared camera \citepads{1998SPIE.3354..606L}.
We obtained a series of $60 \times 3$\,s exposures using a $K_s$ broadband filter\footnote{\url{http://www.eso.org/sci/facilities/paranal/instruments/naco/}}, at mean MJD=57474.4049.
The wavefront sensor was tracking $\alpha$\,Cen~B. The airmass was 1.42 and the visible seeing $1.6\arcsec$ according to the observatory's seeing monitor (DIMM).
We selected the S27 camera, which has a field of view of $27.7\arcsec$. 
The astrometric calibration for the S27 camera is taken from \citetads{2006A&A...456.1165C}: the plate scale is $27.01 \pm 0.05 $\,mas/pix, and the position angle is $+0.00 \pm 0.20$\,degrees. These values are stable over several years within their stated errors, and they are in agreement with the calibrations of \citetads{2003A&A...411..157M} and \citetads{2015A&A...573A.127C}.
We kept $\alpha$\,Cen outside of the NACO field of view to reduce the bleeding on the detector and allow for relatively long exposures.
Due to the complex interaction of the wings and speckles of the two point spread functions, $\alpha$\,Cen is not a good candidate for classical coronagraphic techniques (e.g. as available in SPHERE or NACO).

\begin{figure}[]
        \centering
        \includegraphics[width=4.4cm]{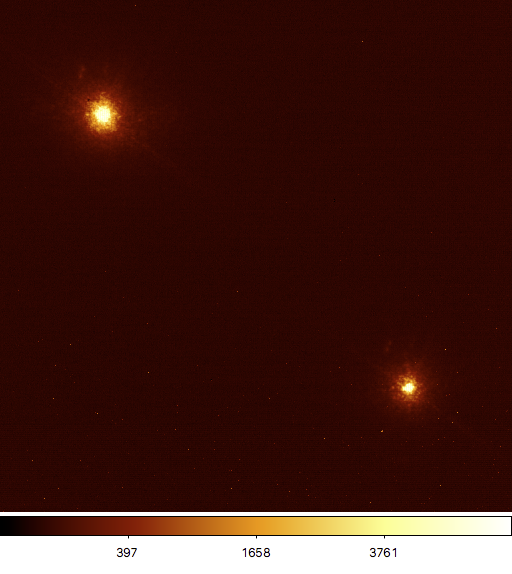}
        \includegraphics[width=4.4cm]{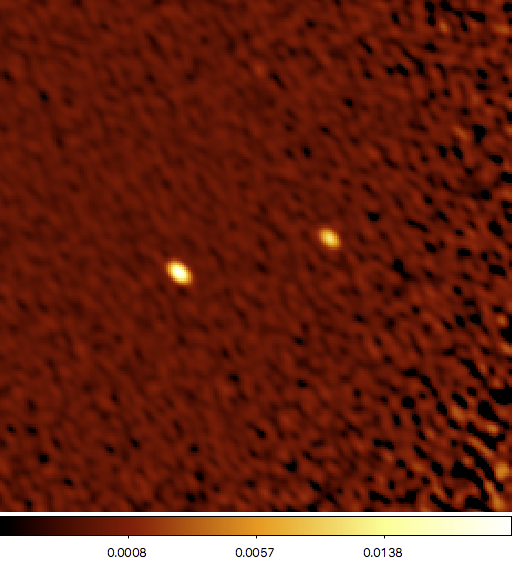}
        \caption{{\it Left panel:} NACO acquisition image of $\alpha$\,Cen A and B obtained on 10 April 2004 (the color scale is in ADU). {\it Right panel:} ALMA image of $\alpha$\,Cen A and B in band 7 (345 GHz) obtained on 7 July 2014 (the color scale is in Jy/beam). The separation of the two stars is $11.12\arcsec$ in the NACO image and $4.24\arcsec$ in the ALMA image. The scale and orientation of the two images is the same (North is up and East to the left).\label{alfcen-naco-alma}}
\end{figure}

In addition to this long exposure, we reduced the associated short exposure acquisition image to derive differential astrometry of components A and B (see Sect.~\ref{astrometryAB}).
We also retrieved from the ESO archive one NACO acquisition image (Fig.~\ref{alfcen-naco-alma}, left panel) obtained on 10 April 2004 (MJD = 53105.1976).
It was taken at an airmass of 1.28 with a DIMM seeing of $0.57\arcsec$ using the S27 camera.
The exposure time was 0.4\,s, using a narrow band filter centered at $1.26\,\mu$m ({\tt NB\_1.26}) and a neutral density filter ({\tt ND\_short}).
The cores of the two star images are saturated but the centroid position can be estimated to an accuracy of half a pixel ($13$\,mas).
These two acquisition images cannot be accurately tied to a global coordinate system,
but they are nevertheless useful as they give a highly accurate relative position of the two stars.

\subsection{ALMA millimeter interferometry}

We took advantage of the observations of $\alpha$\,Cen collected by \citetads{2015A&A...573L...4L} to improve the relatively limited recent astrometry of the $\alpha$\,Cen system.
The accuracy of the absolute referencing of the ALMA coordinate system of these images is limited by the position uncertainty and angular distance of the phase calibrator.
Unfortunately, these observations were not designed to obtain the best possible absolute astrometry, and due to self-referencing of the phases, variable offsets of 1 to $2\arcsec$ with respect to the ICRS are present in the images.
However, the differential astrometry within the interferometric field of view is very accurate, and essentially limited by the signal-to-noise ratio of the images.
This is particularly true for the shorter wavelengths in ALMA bands 7, 8 and 9, where the angular resolution is sufficiently high to obtain a very good relative position accuracy (Fig.~\ref{alfcen-naco-alma}, right panel).
The systematics on the angular separation are linked to the uncertainty on the baseline length and on the effective wavelength of the observations, which are extremely well known.
The position angle is defined by the geometry of the array, also very well known.
As a result, both sources of systematic error are completely negligible compared to the statistical uncertainty of the measurement of the centroid of the star positions in the images.
The relative positions of $\alpha$\,Cen~B with respect to component A are listed in Table~\ref{alfcen-relpos}.

\subsection{Astrometric referencing \label{referencing}}

Due to the extreme brightness of $\alpha$\,Cen, the region around the two stars is poorly covered by astrometric catalogs, and the referencing of the images from the different instruments is a delicate task.
We adopted the approach to consistently reference all our images to the 2MASS coordinate system, that is tied to the ICRS and whose astrometric accuracy is $\pm 0.08\arcsec$ for individual stars brighter than $K_s = 14$ (\citeads{2006AJ....131.1163S}, \citeads{1538-3881-139-6-2440}).
We first matched the \emph{World Coordinate System} (WCS) of the archival NTT/SOFI $J_s$ band image to the 2MASS Atlas $K_s$ band image with reference {\tt ki1260044.fits} retrieved from the All-Sky Release Survey at NASA/IPAC Infrared Science Archive\footnote{\url{http://irsa.ipac.caltech.edu/}}.
The binding of the SOFI image to the 2MASS frame relies on a sample of 14 bright unsaturated stars spread over the SOFI field. These stars were selected to be unblended in the higher resolution SOFI images, and they were identified unambiguously in the 2MASS image.
The standard deviation of the residuals of the star positions with respect to the adjusted SOFI WCS is $0.07\arcsec$. We conservatively adopted an uncertainty of half of this value ($\sigma_\mathrm{SOFI} = 0.035\arcsec$) as our relative astrometric uncertainty in binding the SOFI coordinates to the 2MASS frame.
Following a similar procedure with a set of nine reference stars, we referenced the coordinates of the SUSI2 images from \citetads{2007A&A...464..373K} to the 2MASS frame with a residual standard deviation of $0.11\arcsec$.
We defined the uncertainty in the coordinate referencing as $\sigma_\mathrm{SUSI2} = 0.06\arcsec$.
Finally, using 11 stars detected both on the NACO and SUSI2 images, we set the WCS of our NACO image to the SUSI2 coordinates with a standard deviation of $0.04\arcsec$, giving an uncertainty of $\sigma_\mathrm{NACO-SUSI2} = 0.02\arcsec$.
We preferred to reference the infrared NACO images in the $K_s$ band to the SUSI2 image rather than the SOFI image ($J_s$ band) because the SOFI frame is not as deep as the SUSI2 $I$ band image and therefore shows fewer stars in common. We checked on the brightest stars that the agreement between the coordinate systems of NACO and SOFI is well within $\pm 0.05\arcsec$.
We adopted the same uncertainty on the relative coordinate shift between SOFI and NACO as between SUSI2 and NACO:
$\sigma_\mathrm{NACO-SOFI} = 0.02\arcsec$.

In summary, we obtained $1\sigma$ coordinate accuracies of $\pm 0.035\arcsec$ for the SOFI image with respect to the 2MASS frame and $\pm 0.06\arcsec$ for the NACO image.
As we referenced our positions of $\alpha$\,Cen and the S stars to the SOFI frame, we consider only the difference uncertainty that is $\sigma_\mathrm{NACO-SOFI} = 0.02\arcsec$.


\section{Astrometry of $\alpha$\,Centauri A and B\label{astrometryAB}}

The apparent trajectories of the two components of $\alpha$\,Cen on the sky are defined by the combination of their orbital motion, their parallactic wobble and the proper motion of their center of mass. We explain in this section how we computed these trajectories.

\subsection{Revised orbital elements and parallax\label{orbit}}

\begin{figure*}[ht]
        \centering
        \includegraphics[width=9cm]{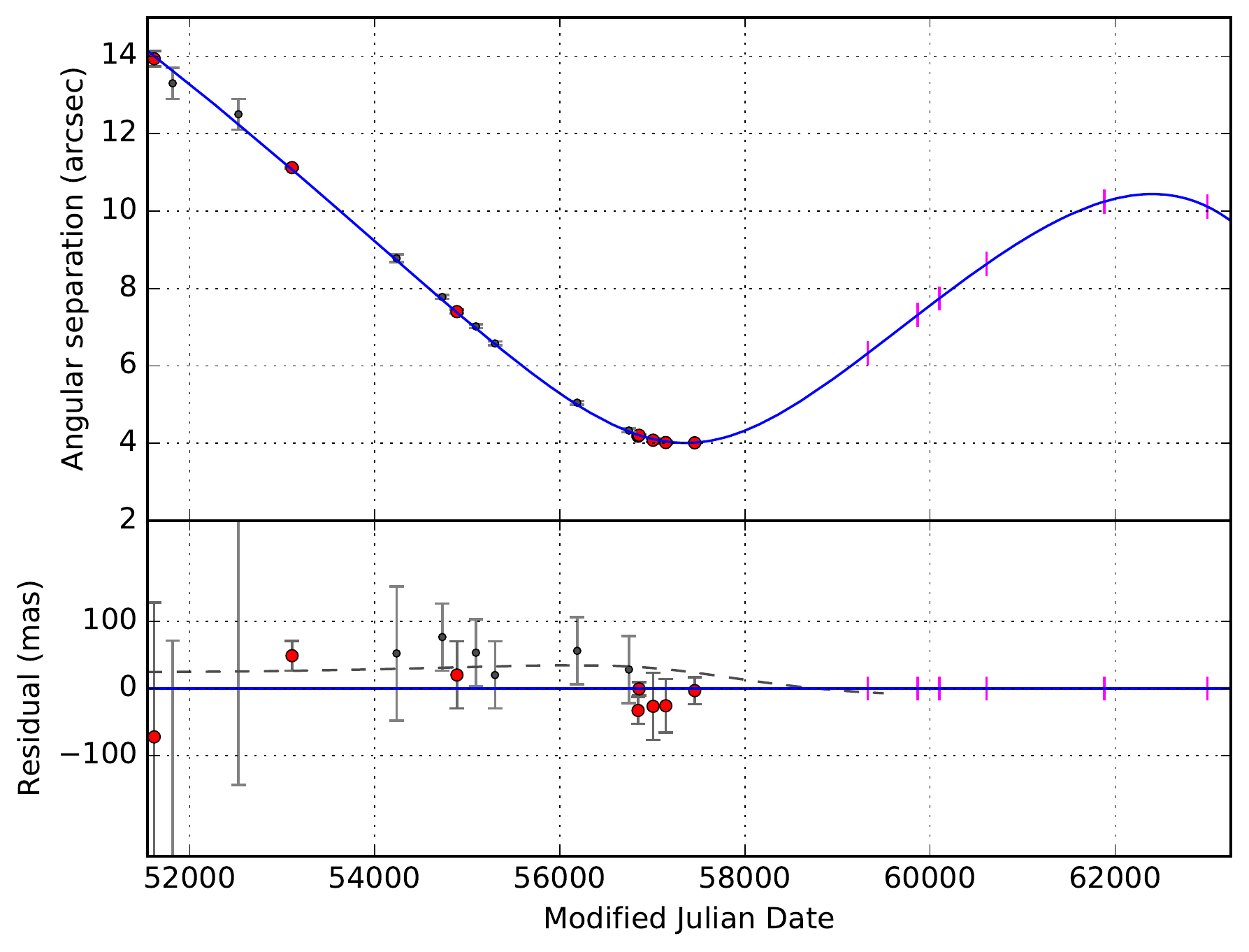}
        \includegraphics[width=9cm]{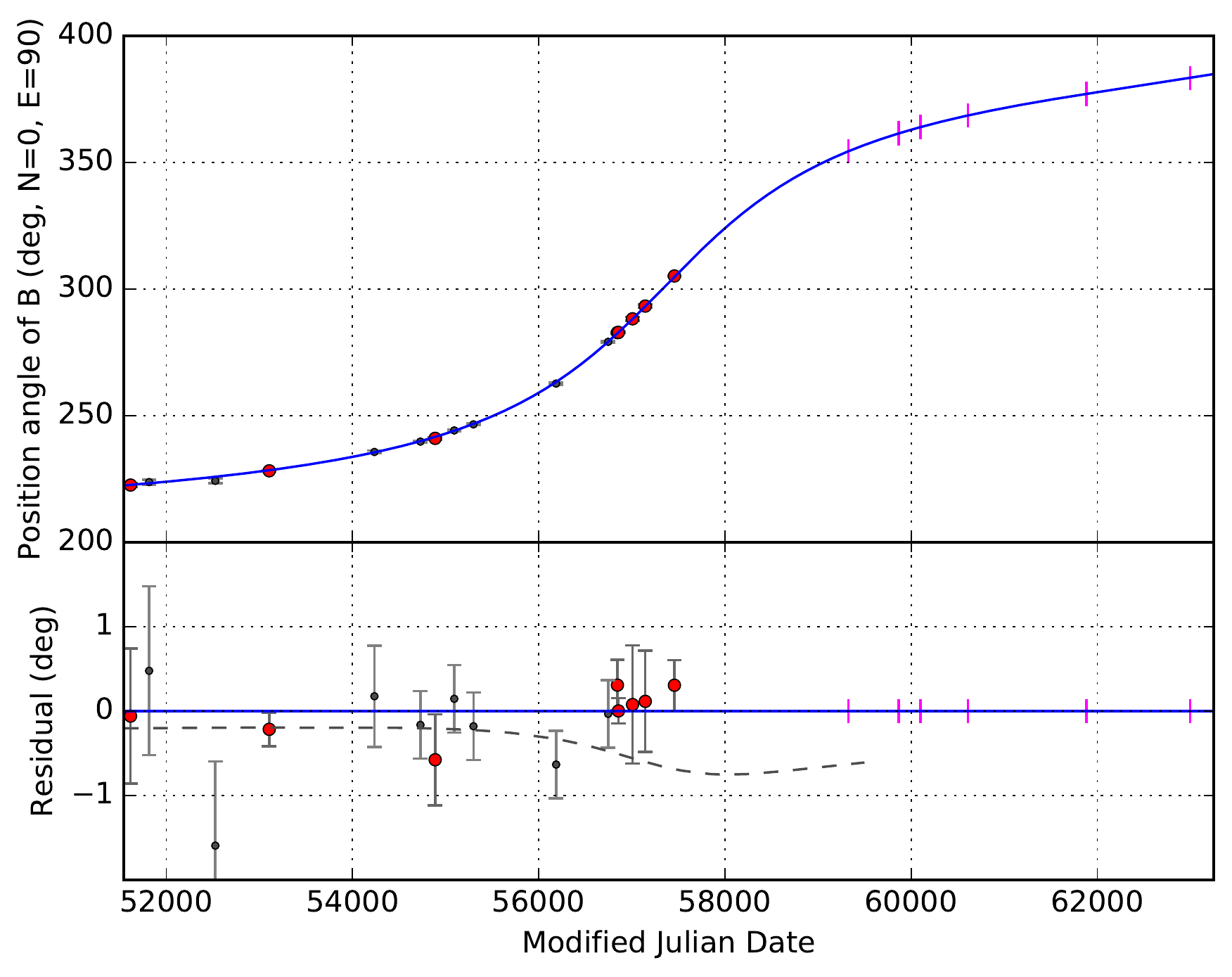}
        \caption{Angular separation $\rho$ (left) and position angle $\theta$ (right) of $\alpha$\,Cen~B relative to $\alpha$\,Cen~A.
        Our new astrometric measurements are shown as red dots, and the literature values from the WDS are shown as gray dots.
        The predictions from our best-fit orbit (Table~\ref{orbit-elements}) are represented with solid blue curves.
        The residuals compared to our best-fit orbit are shown in the bottom panels, together with the differences between the orbit determined by \citetads{2016A&A...586A..90P} and our best-fit orbit  (dashed gray lines, for comparison).
        The time scale runs from 1st January 2000 to 31 December 2031, and the magenta vertical markers indicate the epochs of the conjunctions of S1 to S6.
        \label{orbit-models}}
\end{figure*}

The quality of the orbital elements of $\alpha$\,Cen is essential to predict accurately the positions of the two stars several decades from now.
We therefore compared the predictions in angular separation $\rho$ and position angle $\theta$ obtained using the orbital elements recently published by \citetads{2016A&A...586A..90P} with our new astrometry complemented by archival positions from the literature.
These relative astrometric measurements are listed in Table~\ref{alfcen-relpos}.
The agreement was found to be good at a level of 50\,mas on the separation, but we find a systematically increasing deviation of the position angle $\theta$ from the expected orbit over the last five years, reaching in 2016 approximately $+1^\circ$ (more than $3\sigma$ significance).
This shift is shown with dashed lines in the lower panels of Fig.~\ref{orbit-models}.
This deviation is also visible in the recent CCD camera astrometry by \citetads{2015JDSO...11...81A}.
Our improved astrometric model is the one shown with the blue lines in Fig.~\ref{orbit-models}.

Taking our latest astrometric measurement into account, we thus recomputed the orbital parameters of $\alpha$\,Cen using the same high precision radial velocity data set as \citetads{2016A&A...586A..90P}.
We downloaded the radial velocities from the Ninth Catalog of Spectroscopic Binary Orbits\footnote{\url{http://sb9.astro.ulb.ac.be}} (SB9, \citeads{2004A&A...424..727P}).
The archival astrometric measurements were retrieved from the Washington Double Star catalog (\citeads{2016yCat....102026M}; \citeads{2001AJ....122.3466M}), and the associated uncertainties were taken from the original articles (when available).
We rejected the measurements with large uncertainties as their weight in the fit was negligible.
We chose to give a dominant weight to the ALMA band 9 astrometry obtained at epoch 2014.5426, to make it our fiducial astrometric point.
This is to our knowledge the most accurate differential measurement of the relative position of $\alpha$\,Cen A and B,
with an unbiased precision of 10\,mas on the separation and $0.15^\circ$ on the position angle.

\begin{figure}[ht]
        \centering
        \includegraphics[width=9cm]{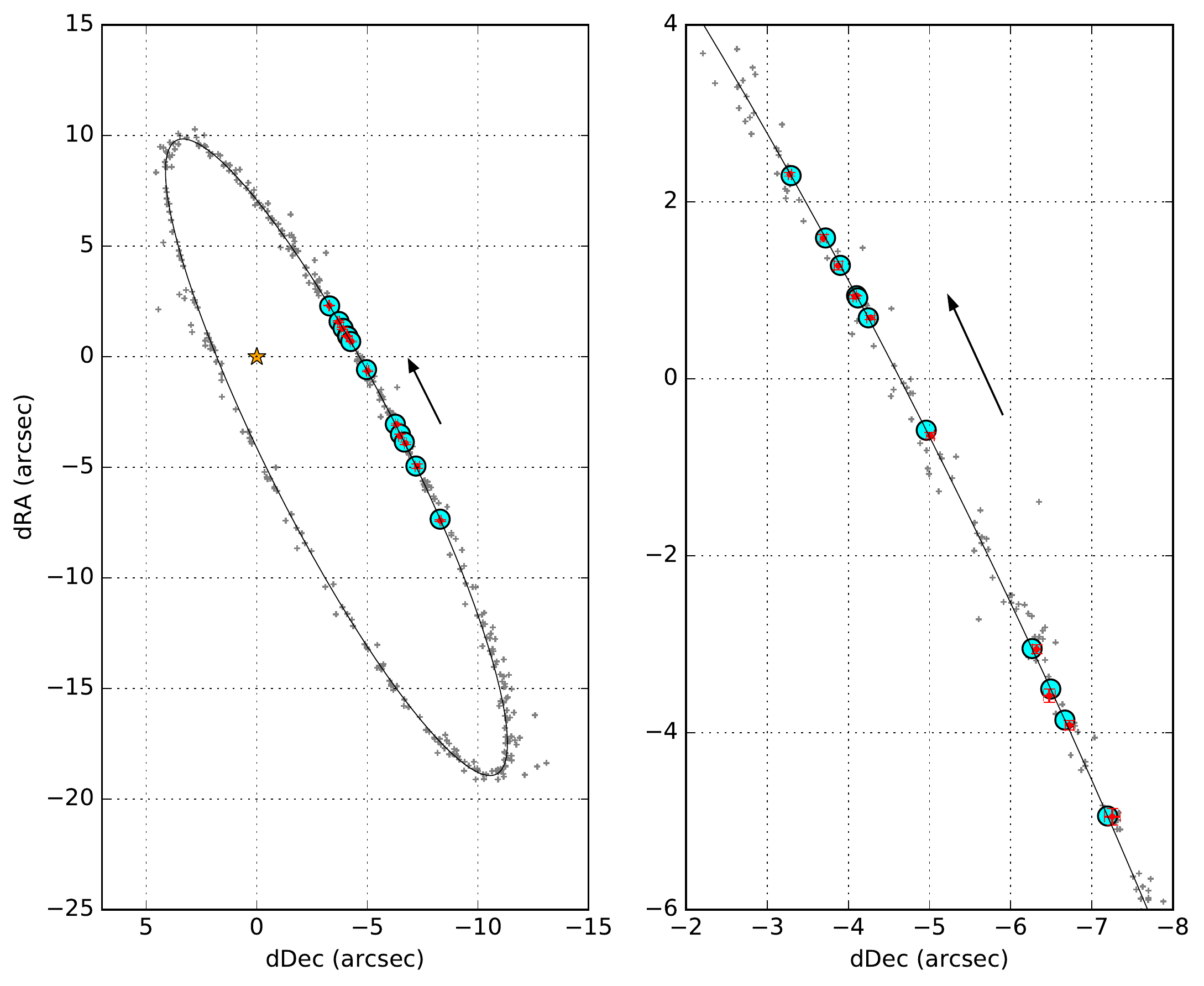}
        \includegraphics[width=9cm]{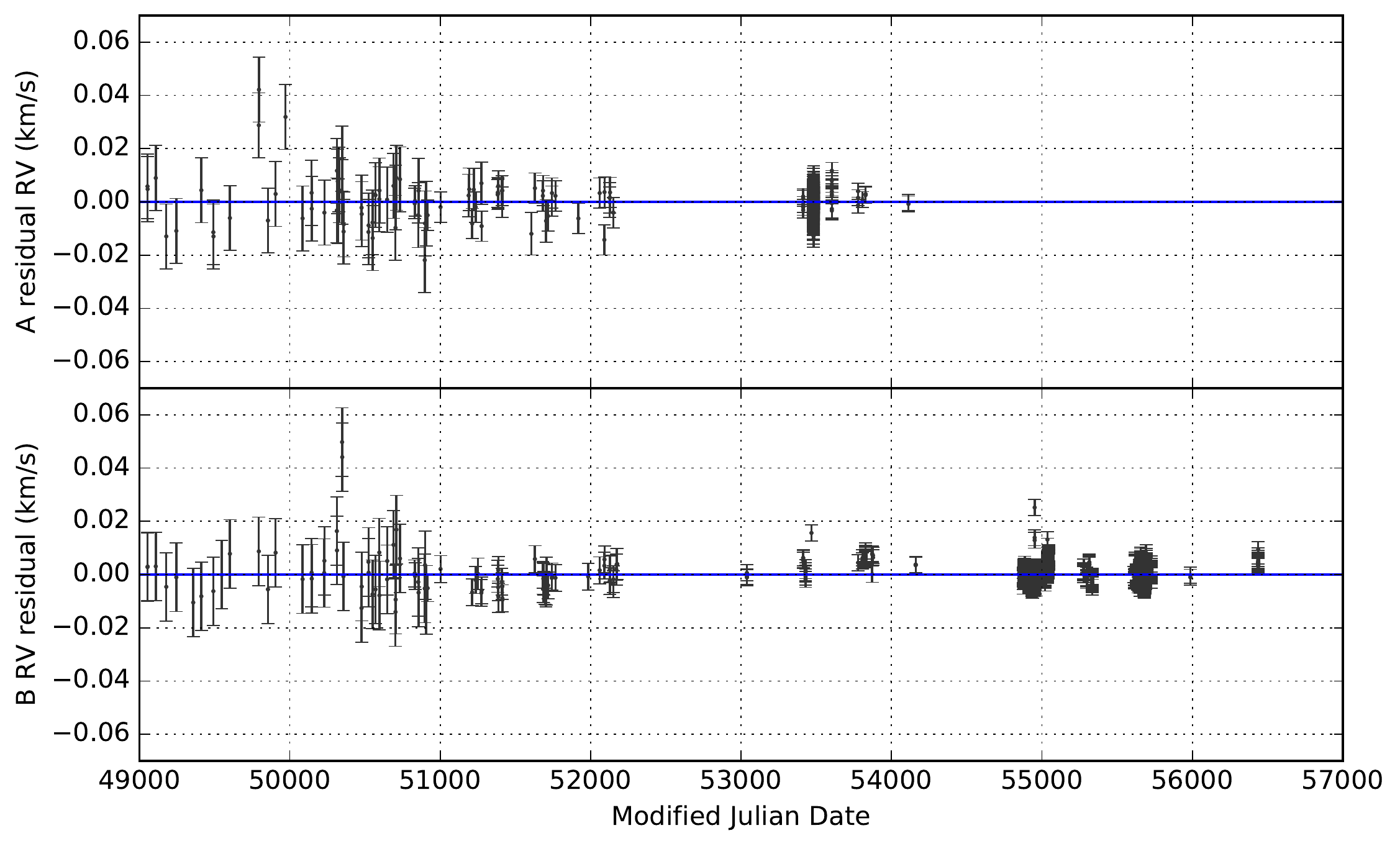}
        \caption{{\it Top:} Adjusted astrometric orbit of $\alpha$\,Cen B relative to $\alpha$\,Cen~A. The top right panel shows an enlargement of measurements since 2002.
        The predicted orbital positions are shown with cyan disks, and the new measurements with red points.
        {\it Bottom:} Residual of the radial velocities of $\alpha$\,Cen A and B with respect to the adjusted orbit. 
        \label{alfcen-orbit}}
\end{figure}

The calculated, projected orbital trajectory of $\alpha$\,Cen~B relative to $\alpha$\,Cen~A is shown in Fig.~\ref{alfcen-orbit}, together with the corresponding radial velocity residuals for the two stars.
Only the recent CES and HARPS radial velocity data are shown in the plot of the residuals, but all the available measurements were included in the fit.
Following \citetads{2016A&A...586A..90P}, we added constant velocity terms $\Delta V_A$ and $\Delta V_B$ for $\alpha$\,Cen A and B to account for their gravitational redshift and convective shift.
\citetads{2016A&A...586A..90P} considered only a shift for component B, but we preferred to consider both components separately as otherwise we observe a systematic residual of the HARPS velocities of A with respect to the best fit orbit.
We obtained a very comparable number as \citetads{2016A&A...586A..90P} for $\alpha$\,Cen~B ($\Delta V_B = 322 \pm 5$\ m\ s$^{-1}$), and a small shift of $\Delta V_A = 8 \pm 5$\ m\ s$^{-1}$ for component A.
We also corrected the zero point of the CES spectrograph with respect to HARPS by $-70$\ m\ s$^{-1}$ for $\alpha$\,Cen~A and $-29$\ m\ s$^{-1}$ for $\alpha$\,Cen~A. This different velocity zero point is not unexpected as the two stars have significantly different spectral types.

Although the uncertainties of the individual HARPS measurements are extremely small (a few 10\ cm\ s$^{-1}$), a much larger dispersion is observed from one measurement to the other.
This dispersion can be either of instrumental of astrophysical origin.
As the two components of $\alpha$\,Cen are close on sky and very bright, contamination from one star on the other can affect the measurements depending on the seeing \citepads{2015IJAsB..14..173B}.
Shifts could also be induced for example by the asteroseismic oscillations of the two stars, or (on a longer timescale) by the presence of star spots.
The rotation periods of $\alpha$\,Cen A and B are of 22 and 41\,days \citepads{2000A&A...363..675M}, and although this corresponds to a relatively slow rotation for a field star, the evolving star spots on the two stars will affect the radial velocities through the Rossiter-McLaughlin effect.
For this reason, we added quadratically a uniform uncertainty of $\sigma = 3$\,m\ s$^{-1}$ to all radial velocity measurement errors, including the HARPS measurements.
This corresponds to the observed dispersion of the HARPS points around the best fit orbit.
This operation does not change significantly the derived parameters, but it results in a more balanced weight to the different observables (astrometry and radial velocities) and produces a more stable fit.

The resulting orbital parameters, parallax and masses of $\alpha$\,Cen are listed in Table~\ref{orbit-elements}.
The overall quality of the fit of the astrometric positions can be assessed from Fig.~\ref{orbit-models}, which compares the model to the observations.
The best-fit reduced $\chi^2$ is 0.95, confirming that the adjusted orbit is a faithful representation of the observations.
Most of the orbital elements are statistically compatible with the values found by \citetads{2016A&A...586A..90P}, but we find a smaller semi-major axis $a$ (this parameter is well constrained by our new astrometry).
The smaller uncertainty on $a$ combined with the high precision radial velocities results in a better defined parallax.
We find $\pi = 747.17 \pm 0.61$\,mas, that is slightly larger than the parallax quoted in \citetads{2016A&A...586A..90P} ($743 \pm 1.3$\,mas) but in line with the one from \citetads{1999A&A...341..121S} ($\pi = 747.1 \pm 1.2$\,mas).
Our parallax is also consistent with the indirect value $\pi = 745.3 \pm 2.5$\,mas predicted by \citetads{2003A&A...404.1087K}
and is almost in between the original \citepads{1997ESASP1200.....E} and revised \citepads{2007A&A...474..653V} \emph{Hipparcos} values.
We obtain significantly lower masses for $\alpha$\,Cen A and B than \citetads{2016A&A...586A..90P}, in particular for component B ($-3.6\%$).
Our values are however in excellent agreement with the previous determinations by \citetads{2002A&A...386..280P}, who adopted the parallax value of \citetads{1999A&A...341..121S}.
We plan to obtain sub-milliarcsecond differential astrometry of $\alpha$\,Cen A and B in the coming years (using e.g.~VLTI/GRAVITY or VLT/NACO) to further improve the orbital elements and masses.

\begin{table}
        \caption{Parallax, proper motion, orbital elements and masses of $\alpha$\,Cen.}
        \centering          
        \label{orbit-elements}
        \begin{tabular}{lrr}
	\hline\hline
        \noalign{\smallskip}
        Parameter & P16$^a$ & K16$^b$ \\
        \noalign{\smallskip}
        \hline    
        \noalign{\smallskip}
	$\pi$ (mas) & $743 \pm 1.3$ & $747.17 \pm 0.61$ \\
        \hline    
        \noalign{\smallskip}
	$\mu_\alpha$ (mas\ yr$^{-1}$)$^c$ & & $\barypmRA \pm \barypmerr$ \\
	$\mu_\delta$ (mas\ yr$^{-1}$)$^c$ & & $\barypmdec \pm \barypmerr$ \\
        \noalign{\smallskip}
	$v_\alpha$ (km\ s$^{-1}$)$^c$ & & $\baryvalpha \pm \baryvalphaerr$ \\
	$v_\delta$ (km\ s$^{-1}$)$^c$ & & $\baryvdelta \pm \baryvdeltaerr$ \\
	$v_r$ (km\ s$^{-1}$)$^c$ &  $-22.390 \pm 0.0042$ & $-22.3930 \pm 0.0043$ \\
        \hline    
        \noalign{\smallskip}
	$a$ ($\arcsec$) & $17.66 \pm 0.026$ & 	$17.592 \pm 0.013$ \\
	$i$ (deg) &  $79.32 \pm 0.044$ & 		$79.320 \pm 0.011$ \\
	$\omega$ (deg) & $232.3 \pm 0.11$ & 	$232.006 \pm 0.051$ \\
	$\Omega$ (deg) & $204.85 \pm 0.084$ & 	$205.064 \pm 0.033$ \\
	$P$ (years) &  $79.91 \pm 0.013$ & 		$79.929 \pm 0.013$ \\
	T$_0$ (year) &  $1955.66 \pm 0.014$ & 	$1955.604 \pm 0.013$ \\
	$e$ &  $0.524 \pm 0.0011$ & 			$0.5208 \pm 0.0011$ \\
	$\Delta V_A$ (m\ s$^{-1}$) &  $-$ & $8 \pm 5$ \\
	$\Delta V_B$ (m\ s$^{-1}$) &  $329 \pm 9$ & $322 \pm 5$ \\
        \hline
        \noalign{\smallskip}
        $m_\mathrm{tot}$ ($M_\odot$) & & $2.0429 \pm 0.0072$ \\
	$m_B$/$m_\mathrm{tot}$ & $0.4617 \pm 0.00044$ & $0.45884 \pm 0.00027$ \\
	$m_A$ ($M_\odot$) & $1.133 \pm 0.0050$ & $1.1055 \pm 0.0039$ \\
	$m_B$ ($M_\odot$) & $0.972 \pm 0.0045$ & $0.9373 \pm 0.0033$ \\
        \hline                      
        \end{tabular}
        \tablefoot{
        $^a$ \citetads{2016A&A...586A..90P}.
        $^b$ Present work.
         $^c$ Barycentric proper motion, tangential velocities and radial velocity.
	}
\end{table}

The excellent accuracy of the derived orbital parameters (Table~\ref{orbit-elements}) allow us to confidently predict the relative position of $\alpha$\,Cen~B with respect to A to an accuracy better than $0.020\arcsec$ at least over the next 100\,years.
We therefore conclude that for the conjunctions that will occur until 2050, the uncertainty on the orbital displacement of $\alpha$\,Cen A and B has a negligible contribution to the astrometric error budget.

\subsection{Apparent trajectories of $\alpha$\,Cen A \& B\label{ephemeris}}

\begin{table*}
        \caption{Positions of $\alpha$\,Cen A and B in the ICRS coordinate frame.}
        \centering          
        \label{alfcen-positions}
        \begin{tabular}{lcclllll}
	\hline\hline
        \noalign{\smallskip}
        Date & Band & MJD$^a$ & Star & RA & Dec & $\sigma [\arcsec]^b$ & Instrument \\
         \hline         
        \noalign{\smallskip}
1991-04-02 & $H_p$ & 48348.250 & $\alpha$\,Cen A & 14:39:40.8985 & -60:50:06.530 & $0.001$ & \emph{Hipparcos}$^c$\\
2000-03-12 & $K_s$ & 51615.383 & $\alpha$\,Cen A & 14:39:36.394 & -60:50:03.31 & $0.20$ & 2MASS$^d$\\
2009-02-25 & $J_s$ & 54887.356 & $\alpha$\,Cen A & 14:39:31.813 & -60:50:00.01 & $0.07$ & SOFI\\
         \hline         
        \noalign{\smallskip}
1991-04-02 & $H_p$ & 48348.250 & $\alpha$\,Cen B & 14:39:39.3908 & -60:50:22.097 & $0.046$ & \emph{Hipparcos}$^c$\\
2000-03-12 & $K_s$ & 51615.383 & $\alpha$\,Cen B & 14:39:35.103 & -60:50:13.56 & $0.20$ & 2MASS$^d$\\
2009-02-25 & $J_s$ & 54887.356 & $\alpha$\,Cen B & 14:39:30.919 & -60:50:03.48 & $0.07$ & SOFI\\
        \hline
        \end{tabular}
        \tablefoot{$^a$ MJD is the average modified julian date of the measurement.
        $^b$ $\sigma$ is the estimated position uncertainty in arcseconds.
	$^c$ Positions taken from the original \emph{Hipparcos} catalog \citepads{1997ESASP1200.....E}.
        $^d$ The listed 2MASS coordinates are measured from the Atlas image in the $K_s$ band.
        }
\end{table*}

\begin{figure*}[]
        \centering
        \includegraphics[width=\hsize]{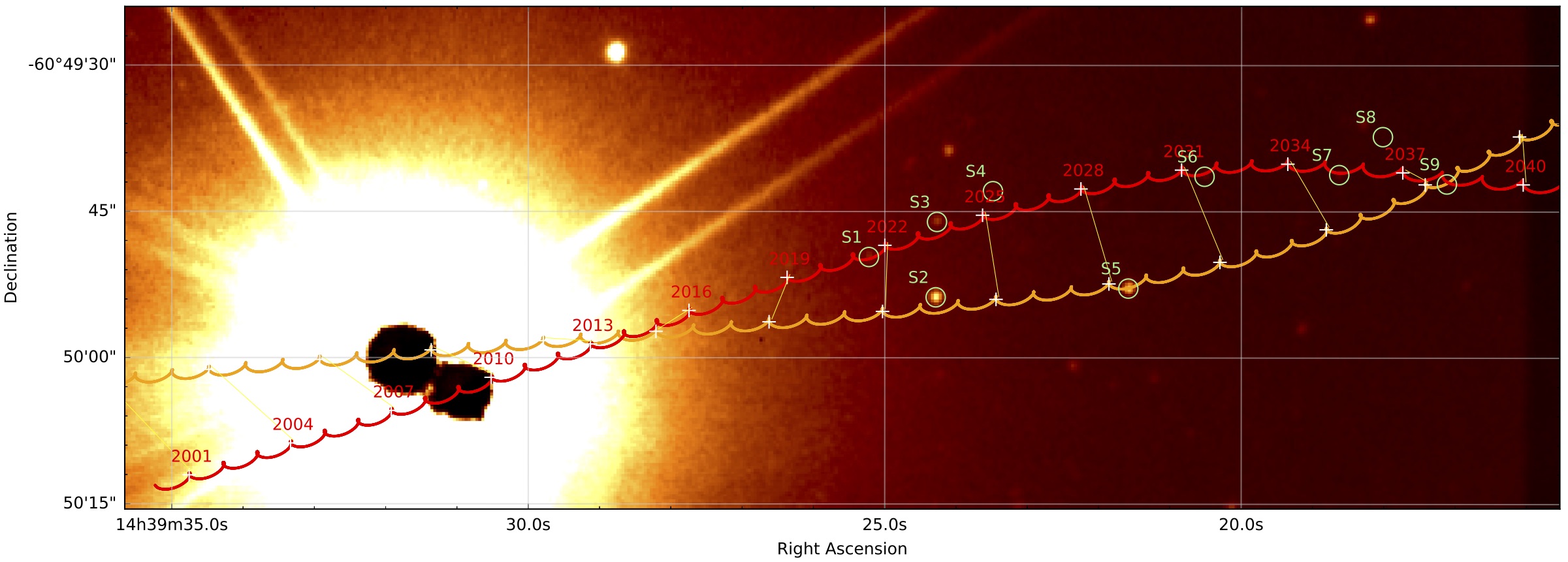}
        \caption{Apparent trajectories of $\alpha$\,Cen A (orange curve) and B (red curve) superimposed on the NTT/SOFI image in the $J_s$ band obtained on 25 February 2009.
        The positions of the approached S stars are circled in green (Sect.~\ref{conjunction-list}).
        \label{sofi-trajectory}}
\end{figure*}

To determine the barycentric proper motion vector of $\alpha$\,Cen, we used the very accurate positions of $\alpha$\,Cen~A and B measured by \emph{Hipparcos} at epoch 1991.25 coupled to the positions of $\alpha$\,Cen A and B we measured in the SOFI image at epoch 2009.1517 (Fig.~\ref{sofi-trajectory}).
We set the masses of the two stars to the values we determined in Sect~\ref{orbit} to compute the position of the barycenter of the system.
The SOFI image provides a robust link between $\alpha$\,Cen and the S star coordinates, as the positions of both $\alpha$\,Cen and the S stars can be measured in the same image.
We therefore took the SOFI image coordinates as our reference point for the barycenter position (Table~\ref{alfcen-positions}).
In spite of the saturation of the images of $\alpha$\,Cen A and B, their individual positions can be measured in the image to $\pm 0.07\arcsec$ (1/4 of the $0.288\arcsec$ pixel) thanks to the long and thin diffraction spikes that are detectable almost to the edge of the image.
We determined the position of their barycenter to an absolute accuracy of $\pm 0.07\arcsec$.
This slightly conservative error estimate also accounts for the absolute referencing
uncertainty of the SOFI coordinates to the 2MASS frame ($\sigma_\mathrm{SOFI} = 0.035\arcsec$).
From the combination of the \emph{Hipparcos} and SOFI barycentric coordinates corrected for parallax, which are separated by almost 19 years, we obtain a measurement of the equatorial coordinate components of the barycentric proper motion vector: $\mu_\alpha = \barypmRA \pm \barypmerr$\,mas\ yr$^{-1}$ and $\mu_\delta = \barypmdec \pm \barypmerr$\,mas\ yr$^{-1}$.
Combining the barycentric proper motion and the parallax, we derived the tangential components of the barycentric velocity: $v_\alpha = \baryvalpha \pm \baryvalphaerr$\,km\ s$^{-1}$ and $v_\delta = \baryvdelta \pm \baryvdeltaerr$\,km\ s$^{-1}$.
The radial component is $v_r = -22.3930 \pm 0.0043$\,km\ s$^{-1}$.
As a consistency check, we observed that the ephemeris positions of components A and B match the observed positions with SOFI (Table~\ref{alfcen-positions}) to $\residualshiftA\arcsec$ and $\residualshiftB\arcsec$, respectively.
 These offsets are smaller than the $\pm 0.07\arcsec$ uncertainty on the measurement of the SOFI positions of A and B.
The relative position of components A and B is therefore in agreement with the predicted orbital position, which gives confidence in the accuracy of our position measurements in the SOFI images. 
The barycenter position is propagated in time using the 3-D velocity vector components to allow for the full 3-D rectilinear motion.
The orbital elements determined in Sect.~\ref{orbit} provide the $\alpha$\,Cen~A-B separation vector as a function of time.
We eventually obtained the celestial coordinates of A and B from the barycenter coordinates and the masses with a local inverse gnomonic projection centered at the barycenter direction.
The resulting trajectories of the two stars are represented in Figs.~\ref{susi2-field}, \ref{alfcen-naco} and \ref{sofi-trajectory}, as well as for the individual S1, S2 and S5 star conjunctions (see Sect~\ref{S1S2S5}).

\section{Near stellar conjunctions with $\alpha$\,Cen A \& B\label{conjunctions}}

\subsection{List of close approaches until 2050\label{conjunction-list}}

The identified close approaches of $\alpha$\,Cen A and B with background sources are listed in Table~\ref{alfcen-approaches}.
We denote in this table the approached stars as ``S stars'' in order of time of closest approach.
The contrast between the approached star and the corresponding $\alpha$\,Cen component was computed using the following magnitudes for $\alpha$\,Cen~A and B: $m_V(A) = 0.01$, $m_V(B)=1.33$, $m_I(A)=-0.68$, $m_I(B)=0.45$, $m_K(A)=-1.49$, $m_K(B)=-0.60$ (the $V-I$ colors of $\alpha$\,Cen were taken from \citeads{2007A&A...474..653V}).
Due to the high density of stars in the plane of the Galaxy, many close conjunctions are detected with very faint stars.
We limited our selection to impact parameters smaller than $5\arcsec$ (6.5\,AU at the distance of $\alpha$\,Cen), with stars brighter than magnitude 16 in the visible ($VRIZ$) and 15 in the infrared ($JHK$ bands).
This angular distance corresponds to the accessible field for narrow-angle astrometry with the VLTI/GRAVITY instrument, and approximately one quarter of the maximum apparent separation between $\alpha$\,Cen A and B.
Considering the capabilities of the current and coming generation of high contrast imaging instrumentation, the selected magnitude limits correspond to a contrast level with the two components of $\alpha$\,Cen that is potentially accessible down to sub-arcsecond angular separations.
We define the minimum approach distance $\rho_\mathrm{min}$ as the geometrical approach in absence of gravitational lensing. In most cases it is very close to the lensed minimum distance, but not for the closest angular separations (within $\approx 0.1\arcsec$) below which the relativistic deflection of the background source light becomes significant (see Sect.~\ref{deflection} for details).
We do not have proper motion measurements for the S stars that will be approached after 2034 (starting with S7, see Table~\ref{alfcen-approaches}).  To estimate the corresponding uncertainty on $\rho_\mathrm{min}$, we considered an arbitrary uncertainty on their proper motion of 10\,mas\ yr$^{-1}$ that was scaled considering the date at which their position was last measured and the predicted date of closest approach. Such a proper motion corresponds to a sky velocity of 20\,km\ s$^{-1}$ at a distance of 400\,pc, and is therefore a reasonable statistical upper limit for the faint and probably distant S stars. The resulting uncertainty was added quadratically to the error bar of $\rho_\mathrm{min}$ and generally dominates the total error budget on this parameter.
This uncertainty will be reduced in the future, as \emph{Gaia} will provide accurate positions and proper motions for the S stars.

As the proper motion of $\alpha$\,Cen in sky coordinates is approximately $3.685$\,arcsec\ yr$^{-1}$ (Table~\ref{orbit-elements}), the uncertainty on the dates of the closest approaches is $\pm 10$\,days for a typical position uncertainty of the S stars of $0.1\arcsec$.

\begin{sidewaystable*}
        \caption{Close approaches within $\rho = 5\arcsec$ of $\alpha$\,Cen A or B between 1st January 2016 and 25 December 2049 for background stars brighter than magnitude $m=16$ at visible wavelengths ($VRIZ$) and $m=15$ in the near-infrared ($JHK$).
        The listed S star coordinates (ICRS) correspond to the indicated measurement epoch (no proper motion correction is included).
        The contrast between the corresponding $\alpha$\,Cen star is listed in the $\Delta V$, $\Delta I$ and $\Delta K$ columns.
        The error bars on the photometry are given in subscript of the corresponding values.}
         \centering          
        \label{alfcen-approaches}
        \begin{tabular}{lcclccccccccccrrr}
	\hline\hline
        \noalign{\smallskip}
        $\star^a$ & $\rho_\mathrm{min}[\arcsec]$ & Date$^b$ & \#$^c$ & K07$^d$ & $m_V$ & $\Delta V$ & $m_I$ & $\Delta I$ & $m_K$ & $\Delta K$ & RA & Dec & Epoch & $\mu_\alpha$$^e$ & $\mu_\delta$$^d$ & $\sigma_\mu^d$ \\
         \hline         
        \noalign{\smallskip}
B & $\minSone \pm \errminSone$ & \dateSone & S1 & 1266 & $20.0_{ 0.5}$ & $18.7$ & $17.6_{ 0.2}$ & $17.2$  & $13.58_{ 0.17}$ & $14.2$ & 14:39:25.224 & -60:49:49.67 & 2009.152 & \PMRASone & \PMdecSone & \PMerrSone \\
A & $\minStwo \pm \errminStwo$ & \dateStwo & S2 & 1184 & $15.7_{ 0.2}$ & $15.7$ & $13.3_{ 0.2}$ & $13.9$  & $11.14_{ 0.03}$ & $12.6$ & 14:39:24.283 & -60:49:53.80 & 2016.236 &  \PMRAStwo & \PMdecStwo & \PMerrStwo \\
B & $\minSthree \pm \errminSthree$ & \dateSthree & S3 & 1181 & $19.7_{ 0.3}$ & $18.4$ & $16.0_{ 0.2}$ & $15.6$  & $12.89_{ 0.11}$ & $13.5$ & 14:39:24.264 & -60:49:46.03 & 2009.152 &  \PMRASthree & \PMdecSthree & \PMerrSthree  \\
B & $\minSfour \pm \errminSfour$ &\dateSfour & S4 & 1113 & $17.5_{ 0.2}$ & $16.2$ & $15.9_{ 0.2}$ & $15.5$  & $-$ & $-$ & 14:39:23.485 & -60:49:42.91 & 2016.236 &  \PMRASfour & \PMdecSfour & \PMerrSfour \\
A & $\minSfive \pm \errminSfive$ & \dateSfive & S5 & 0951 & $21.5_{0.9}$ & $21.5$ & $18.7_{ 0.3}$ & $19.3$  & $ 7.76_{0.02}$ & $9.3$ & 14:39:21.586 & -60:49:52.89 & 2016.236 &  \PMRASfive & \PMdecSfive & \PMerrSfive \\
 B & $\minSsix \pm \errminSsix$ & \dateSsix & S6 & 0856 & $16.9_{0.2}$ & $15.6$ & $15.4_{0.2}$ & $14.9$  & $-$ & $-$ & 14:39:20.518 & -60:49:41.41 & 2009.152 &  \PMRASsix & \PMdecSsix & \PMerrSsix \\
 \hline
        \noalign{\smallskip}
 B & $0.194 \pm 0.318$ & 2034-11-20 & S7 & 0717 & $17.1_{0.2}$ & $15.8$ & $15.3_{0.2}$ & $14.9$ & $-$ & $-$ & 14:39:18.629 & -60:49:41.23 & 2004.252 & $-$ & $-$ & $-$ \\
 B & $3.417 \pm 0.338$ & 2036-11-28 & S8 & 0674 & $22.9_{1.0}$ & $21.5$ & $19.2_{0.3}$ & $18.8$ & $12.84_{0.08}$ & $13.4$ & 14:39:18.017 & -60:49:37.33 & 2004.252 & $-$ & $-$ & $-$ \\
 B & $0.216 \pm 0.342$ & 2037-05-28 & S9 & 0601 & $22.2_{1.0}$ & $20.9$ & $19.9_{0.4}$ & $19.5$ & $11.91_{0.05}$ & $12.5$ & 14:39:17.119 & -60:49:42.18 & 2004.252 & $-$ & $-$ & $-$ \\
 B & $0.487 \pm 0.381$ & 2041-04-29 & S10 & 0463 & $-$ & $-$ & $-$ & $-$ & $10.67_{0.04}$ & $11.3$ & 14:39:15.142 & -60:49:43.66 & 2004.252 & $-$ & $-$ & $-$ \\
 A & $0.263 \pm 0.385$ & 2041-10-15 & S11 & 0462 & $17.6_{0.2}$ & $17.6$ & $15.8_{0.2}$ & $16.5$ & $-$ & $-$ & 14:39:15.107 & -60:49:34.88 & 2004.252 & $-$ & $-$ & $-$ \\
 B & $0.752 \pm 0.382$ & 2041-06-09 & S12 & 0438 & $14.6_{0.2}$ & $13.2$ & $13.3_{0.2}$ & $12.8$ & $-$ & $-$ & 14:39:14.753 & -60:49:42.66 & 2004.252 & $-$ & $-$ & $-$ \\
 A & $1.120 \pm 0.395$ & 2042-10-18 & S13 & 0420 & $-$ & $-$ & $-$ & $-$ & $13.17_{0.08}$ & $14.7$ & 14:39:14.508 & -60:49:34.19 & 2004.252 & $-$ & $-$ & $-$ \\
 B & $3.168 \pm 0.420$ & 2045-04-27 & S14 & 0361 & $23.1_{1.0}$ & $21.8$ & $19.1_{0.3}$ & $18.6$ & $13.05_{0.06}$ & $13.7$ & 14:39:13.375 & -60:49:39.35 & 2004.252 & $-$ & $-$ & $-$ \\
 B & $2.600 \pm 0.429$ & 2046-05-02 & S15 & 0313 & $-$ & $-$ & $15.7_{0.2}$ & $15.3$ & $-$ & $-$ & 14:39:12.596 & -60:49:45.89 & 2004.252 & $-$ & $-$ & $-$ \\
 B & $4.261 \pm 0.430$ & 2046-05-16 & S16 & 0296 & $-$ & $-$ & $18.1_{0.2}$ & $17.7$ & $13.02_{0.06}$ & $13.6$ & 14:39:12.262 & -60:49:47.38 & 2004.252 & $-$ & $-$ & $-$ \\
 A & $3.121 \pm 0.454$ & 2048-10-24 & S17 & 0251 & $14.0_{0.2}$ & $14.0$ & $12.6_{0.2}$ & $13.3$ & $11.78_{0.04}$ & $13.3$ & 14:39:11.421 & -60:49:28.24 & 2004.252 & $-$ & $-$ & $-$ \\
 B & $0.269 \pm 0.460$ & 2049-06-12 & S18 & 0213 & $22.8_{1.0}$ & $21.5$ & $18.5_{0.2}$ & $18.1$ & $12.74_{0.09}$ & $13.3$ & 14:39:10.515 & -60:49:41.44 & 2004.252 & $-$ & $-$ & $-$ \\
 \hline
        \end{tabular}
        \tablefoot{$^a$ The $\star$ column lists the component of $\alpha$\,Cen concerned by the conjunction.
        $^b$ The uncertainty on the closest approach date is linearly proportional to $\sigma(\rho_\mathrm{min})$ and equals approximately $\sigma(\mathrm{Date}) = \pm 10$\,days for $\sigma(\rho_\mathrm{min}) = \pm 0.10\arcsec$.
        $^c$ Star labels as used in the text.
        $^d$ K07: reference number of the approached star in the catalog by \citetads{2007yCat..34640373K}.
        $^e$ The proper motion and associated uncertainty is listed in mas\ yr$^{-1}$.}
\end{sidewaystable*}

\subsection{Approaches of S1 to S6 until 2031\label{S1S2S5}}

We discuss here the cases of stars S1 to S6 that will be approached by the components of $\alpha$\,Cen prior to 2031.
Table~\ref{S-photometry} lists their existing photometric measurements and Table~\ref{S-astrometry} gives their positions measured at different epochs, which we used to estimate their proper motions.
None of the proper motions that we obtained is statistically significant.
For this reason, we considered a null proper motion for the computation of the impact parameter $\rho_\mathrm{min}$.
We however scaled the proper motion uncertainty for each star S1 to S6, taking into account the time of the foreseen closest approach and the date at which their position was last measured.
The resulting uncertainty was added quadratically to the impact parameter error bar.
We adjusted to the photometry the spectral energy distribution (SED) models from the library by \citetads{2004astro.ph..5087C}\footnote{\url{http://www.stsci.edu/hst/observatory/crds/castelli_kurucz_atlas.html}} parametrized by the effective temperature $T_\mathrm{eff}$, the limb darkened angular diameter $\theta_\mathrm{LD}$ and the color excess $E(B-V)$.
There is a partial degeneracy between the reddening (computed assuming a classical Galactic extinction law with $R_V = A_V / E(B-V)=3.1$ and the intrinsic color of the source (i.e.,~its effective temperature).
As a consequence, broadband photometry alone is insufficient to determine precisely the nature of the S stars, but we could obtain reasonably good fits of their SEDs.
Taking advantage of the current relatively large angular separation of the S stars with $\alpha$\,Cen, we plan to better characterize the S stars in the future (in particular S5) with spectroscopic observations.
The positions and proper motions of all the approached stars will also be refined by \emph{Gaia}.

\subsubsection{Approach of $\alpha$\,Cen B to Star S1 in May 2021}

\begin{figure*}
        \includegraphics[height=7.5cm]{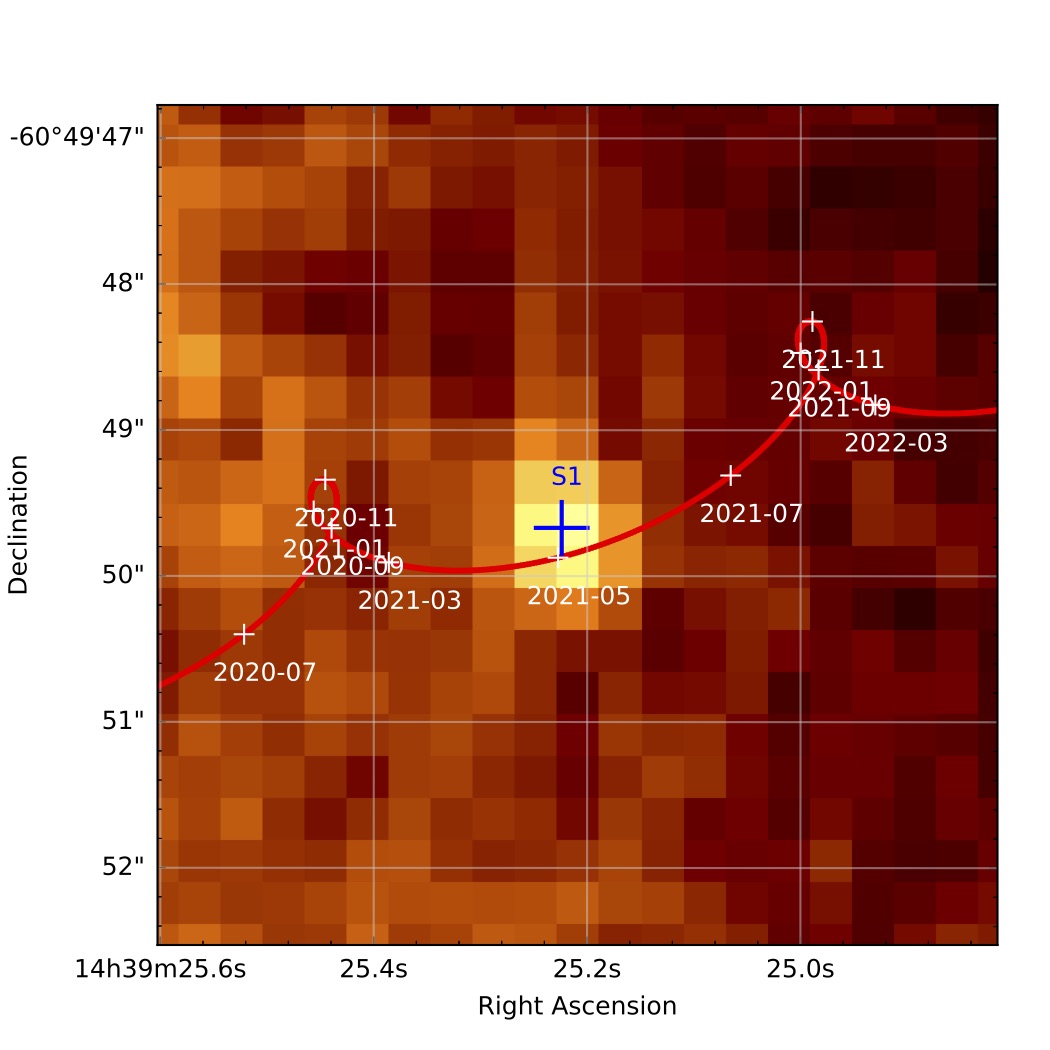}
        \includegraphics[height=7.5cm]{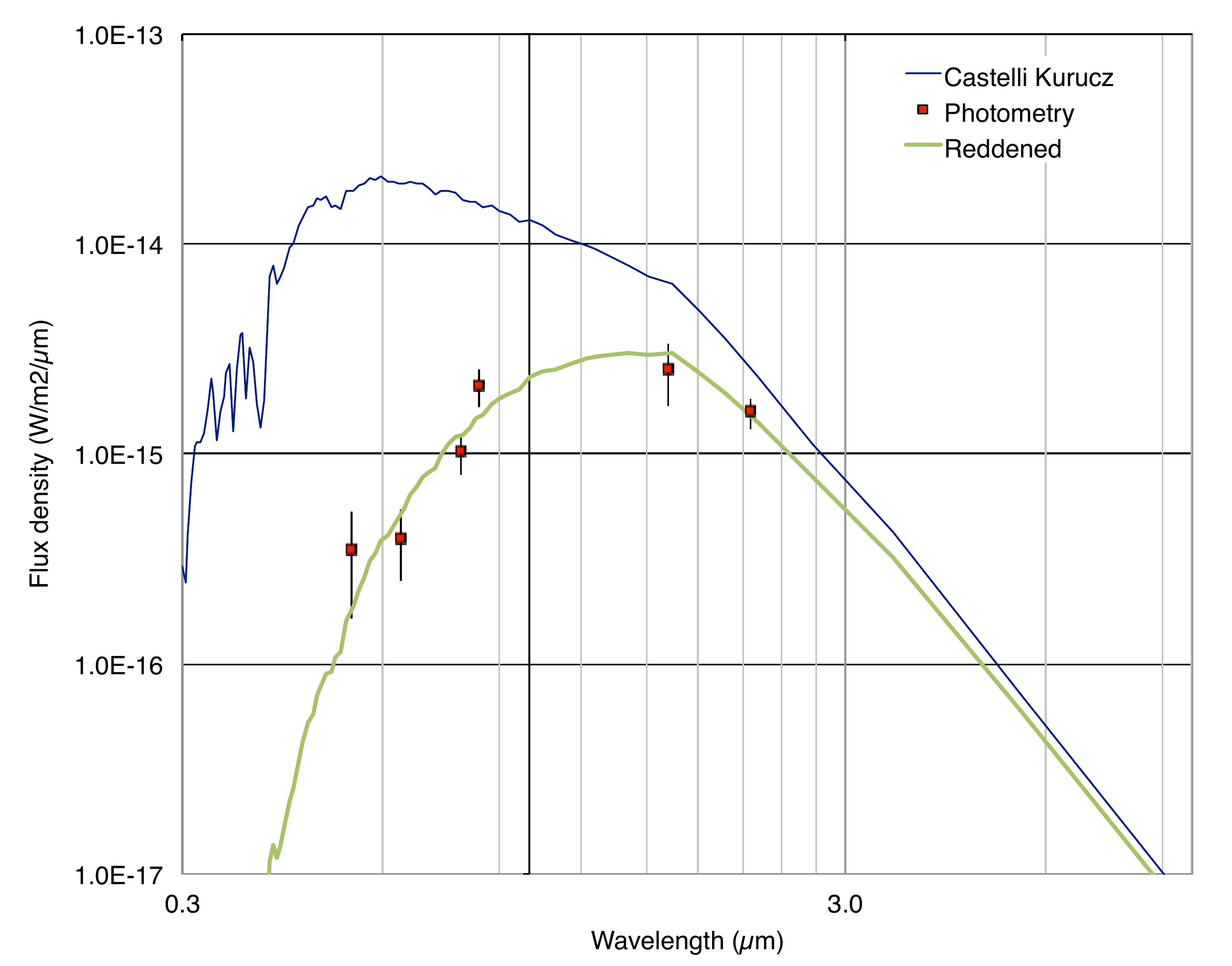}
        \caption{{\it Left panel:} Trajectory of $\alpha$\,Cen B relative to S1 (red curve) on the SOFI image obtained on 25 February 2009. The uncertainty on the impact parameter is represented with blue segments crossing at S1's position. {\it Right panel:} Photometry of S1 (red dots) compared to a Kurucz SED model of a $\theta_\mathrm{LD} = 0.012$\,mas star with $T_\mathrm{eff} = 4500$\,K (blue curve) and $E(B-V)=1.6$ (green curve).
        \label{S1-figures}}
\end{figure*}

Star S1 will be approached by $\alpha$\,Cen~B, with a minimum separation of $\rho_\mathrm{min}(\mathrm{S1}) = \minSone \pm \errminSone\arcsec$ on {\dateSone} $\pm 18$\,days.
Based on astrometry between 2000 and 2009, the derived proper motion of S1 is $\mu_\alpha(\mathrm{S1}) = \PMRASone \pm \PMerrSone$\ mas\,yr$^{-1}$ and $\mu_\delta(\mathrm{S1}) = \PMdecSone \pm \PMerrSone$\,mas\,yr$^{-1}$.
The SED of S1 is poorly constrained as the star is faint at visible wavelengths and infrared photometry beyond the $K$ band is not available.
As shown in Fig.~\ref{S1-figures}, we obtained a correct fit of the SED for an angular diameter of $\theta_\mathrm{LD}(\mathrm{S1}) = 0.012 \pm 0.002$\,mas, an effective temperature of $T_\mathrm{eff}(\mathrm{S2}) = 4500 \pm 1500$\,K and a color excess $E(B-V) = 1.6 \pm 0.5$.

\subsubsection{Approach of $\alpha$\,Cen A to Star S2 in April 2023}

\begin{figure*}
        \includegraphics[height=7.5cm]{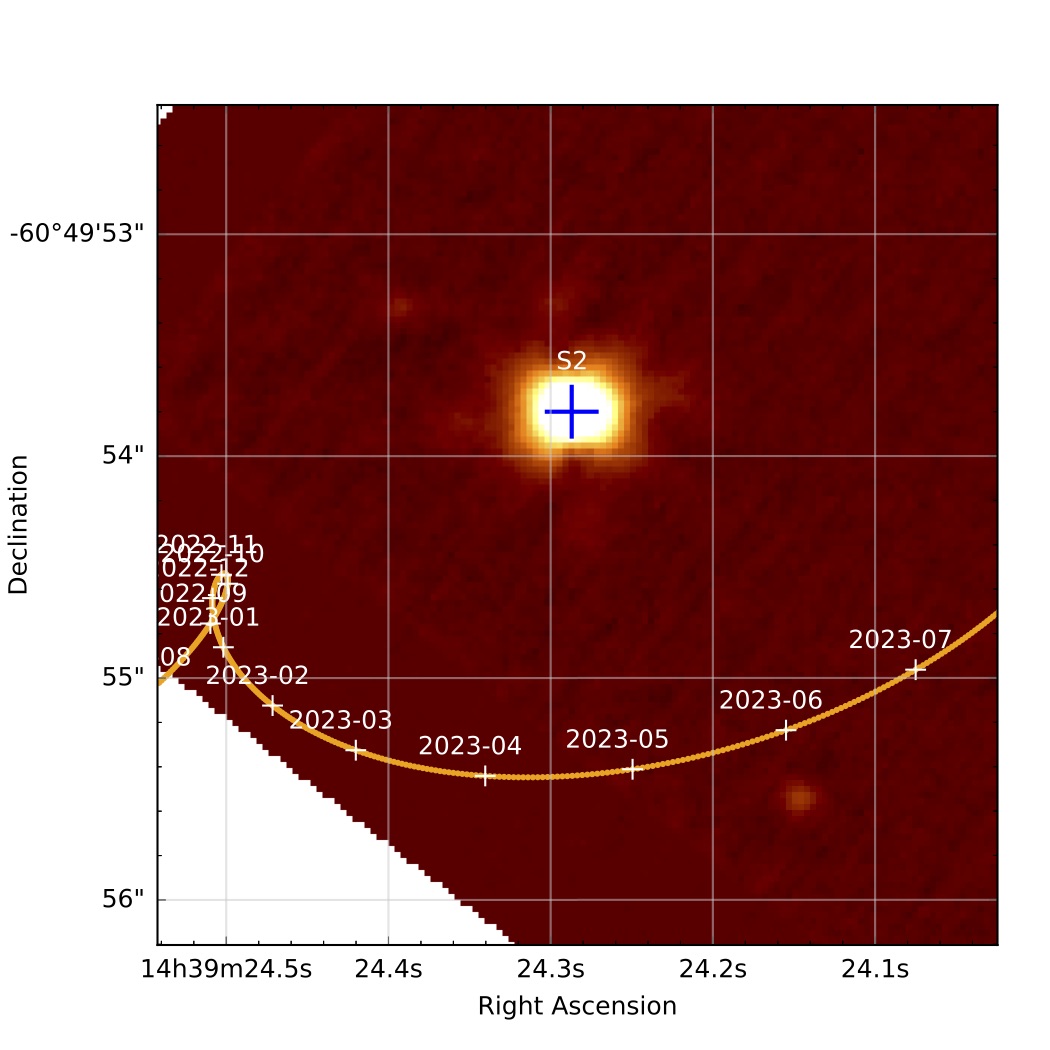}
        \includegraphics[height=7.5cm]{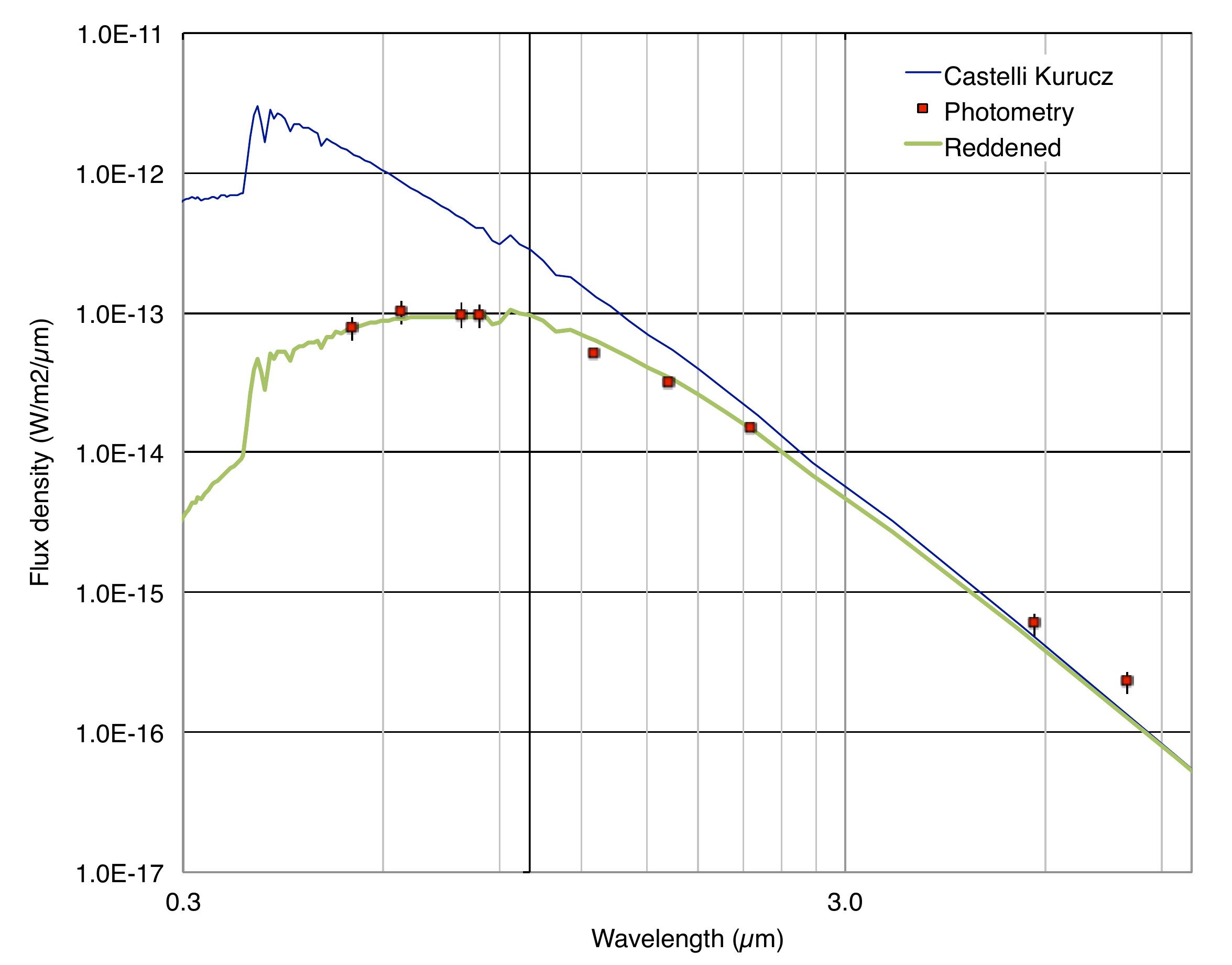}
        \caption{{\it Left panel:} Trajectory of $\alpha$\,Cen A relative to S2 (orange curve) on the NACO image obtained on 27 March 2016. The uncertainty on the impact parameter is represented with blue segments crossing at S2's position. {\it Right panel:} Photometry of S2 (red dots) compared to a Kurucz model with $\theta_\mathrm{LD} = 0.025$\,mas and $T_\mathrm{eff} = 8000$\,K (blue curve) reddened with $E(B-V)=1.0$ (green curve).
        \label{S2-figures}}
\end{figure*}

Star S2 will be approached by $\alpha$\,Cen~A at a minimum separation of $\rho_\mathrm{min}(\mathrm{S2}) = \minStwo \pm \errminStwo\arcsec$ on {\dateStwo} $\pm 11$\,days (Fig.~\ref{S2-figures}, left panel). Based on astrometry spread between 2000 and 2016 (Table~\ref{S-astrometry}), the proper motion of S2 is found to be very slow at $\mu_\alpha(\mathrm{S2}) = \PMRAStwo \pm \PMerrStwo$\ mas\,yr$^{-1}$ and $\mu_\delta(\mathrm{S2}) = \PMdecStwo \pm \PMerrStwo$\ mas\,yr$^{-1}$.

From the fit of its photometry (Fig.~\ref{S2-figures}, right panel) S2 is probably a hot ($T_\mathrm{eff}(\mathrm{S2}) = 8000 \pm 1000$\,K) and relatively nearby star as its reddening is $E(B-V) = 1.0 \pm 0.1$.
The \emph{Spitzer}/IRAC measurements of S2 at 5.8 and $8.0\,\mu$m appear to be affected by the close presence of $\alpha$\,Cen, and although they are represented in Fig.~\ref{S2-figures}, they were not used to constrain the fit.
Its angular diameter of $\theta_\mathrm{LD}(\mathrm{S2}) = 0.025 \pm 0.003$\,mas is compatible with a main sequence A-type star ($R \approx 1.5\,R_\odot$) located at a distance of 550\,pc.

\subsubsection{Approach of $\alpha$\,Cen B to Star S3 in December 2023}

The minimum distance between S3 and $\alpha$\,Cen~B will be reached on {\dateSthree} $\pm 19$\,days at $\rho_\mathrm{min} = \minSthree \pm \errminSthree\arcsec$. The impact parameter is relatively large and the contrast of 13.5 magnitudes in the $K$ band with $\alpha$\,Cen~B is high.
$\alpha$\,Cen will be difficult to observe as it will be relatively close to the Sun at the time of the closest approach.

\begin{figure}
        \includegraphics[width=\hsize]{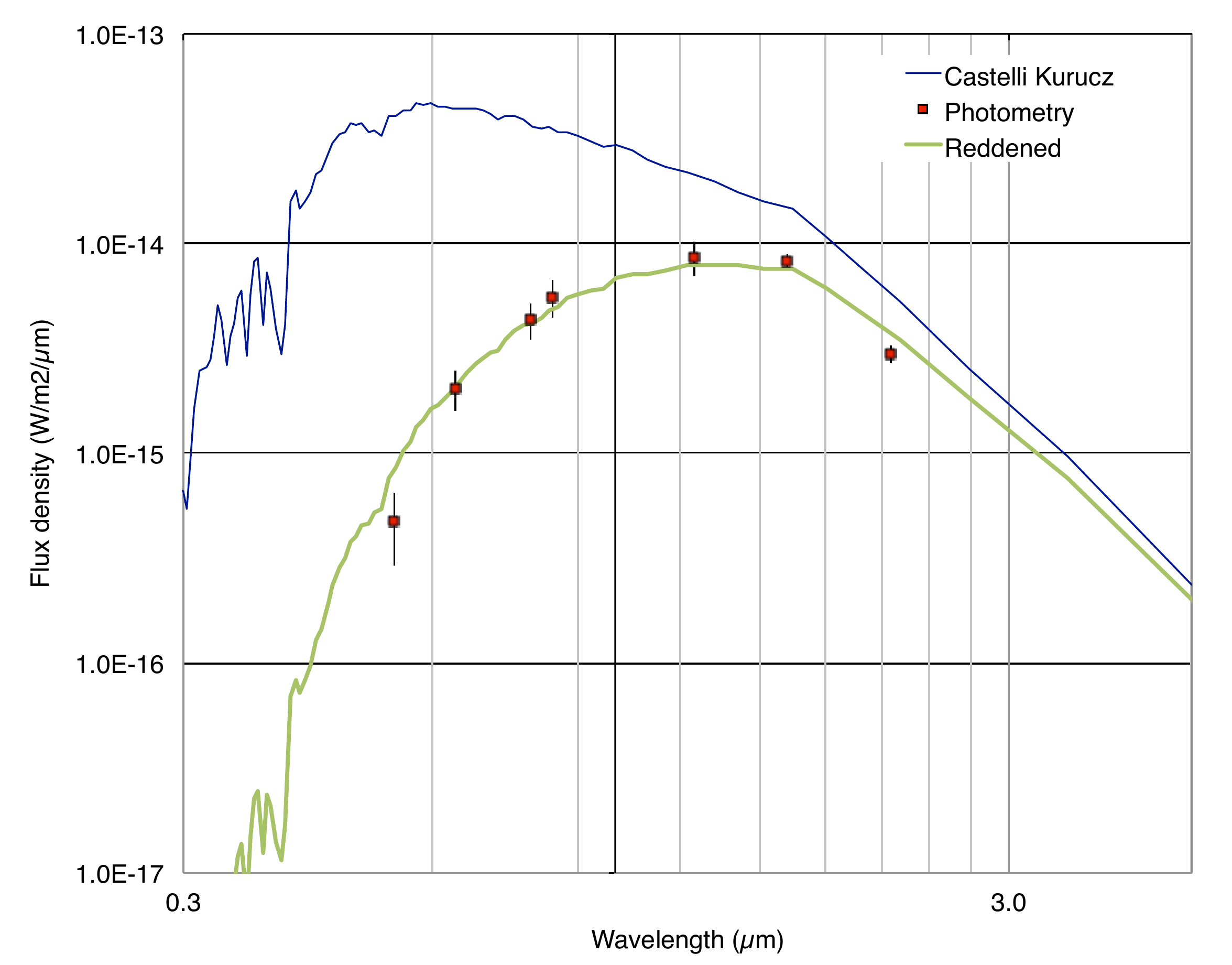}
        \caption{Photometry of S3 (red dots) compared to a Kurucz SED with $\theta_\mathrm{LD} = 0.018$\,mas, $T_\mathrm{eff} = 4500$\,K (blue curve) reddened with $E(B-V)=1.4$ (green curve).
        \label{S3-sed}}
\end{figure}

The SED fit gives an angular diameter $\theta_\mathrm{LD}(\mathrm{S3}) = 0.018 \pm 0.002$\,mas and an effective temperature of $T_\mathrm{eff}(\mathrm{S3}) = 4500 \pm 1000$\,K for $E(B-V)=1.4 \pm 1.0$ (Fig.~\ref{S3-sed}).
It is interesting to remark that S3 was possibly detected by XMM-Newton with an X-ray flux of $1.2 \pm 0.2\ 10^{-13}$\ W\,m$^{-2}$ between 0.2 and 12\,keV \citepads{2013yCat.9044....0X}.

\subsubsection{Approach of $\alpha$\,Cen B to Star S4 in October 2024}

The closest angular distance of S4 and $\alpha$\,Cen~B will be reached on {\dateSfour} $\pm 12$\,days at $\rho_\mathrm{min} = \minSfour \pm \errminSfour\arcsec$.
The impact parameter is large and S4 will unfortunately not be easily observable at the time of the closest approach.
The SED fit is made difficult and uncertain by the absence of infrared photometry. We obtained a probable angular diameter $\theta_\mathrm{LD}(\mathrm{S4}) = 0.008 \pm 0.003$\,mas and an effective temperature of $T_\mathrm{eff}(\mathrm{S4}) = 4500 \pm 1500$\,K for $E(B-V) = 0.3 \pm 0.3$. The low (but poorly constrained) reddening and (marginally) significant proper motion would be compatible with a relatively nearby dwarf star.

\subsubsection{Approach of $\alpha$\,Cen A to Star S5 in May 2028}

\begin{figure*}
        \includegraphics[height=7.5cm]{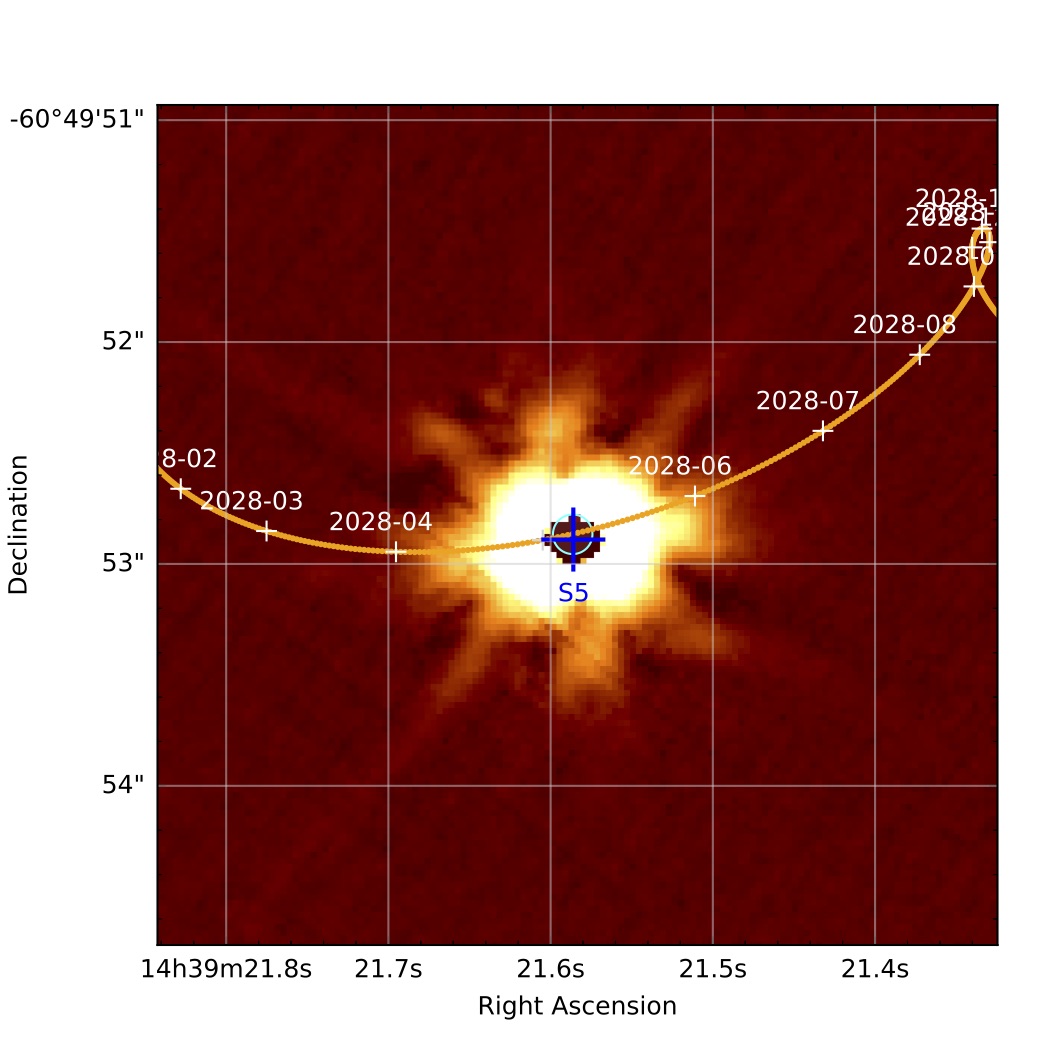}
        \includegraphics[height=7.5cm]{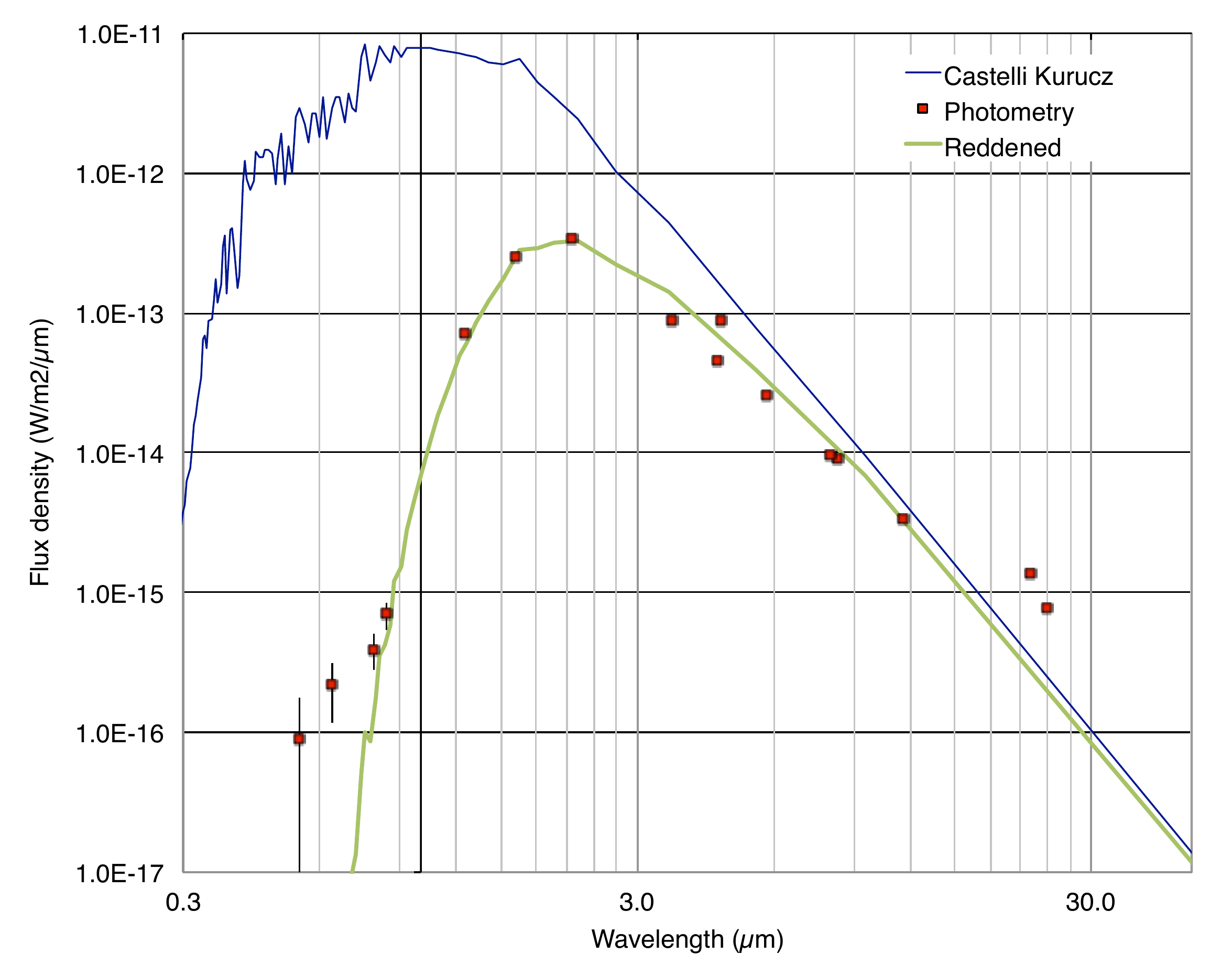}
        \caption{{\it Left panel:} Trajectory of $\alpha$\,Cen A relative to S5 (orange curve) on the NACO image obtained on 27 March 2016. The position of the Einstein ring of $\alpha$\,Cen~A is represented with a light blue circle at the time of the closest approach, and the uncertainty on the impact parameter is represented with blue segments crossing at S5's position.
        {\it Right panel:} Photometry of S5 (red dots) compared to a Kurucz SED with $\theta_\mathrm{LD} = 0.47$\,mas, $T_\mathrm{eff} = 3500$\,K (blue curve) reddened with $E(B-V)=6.5$ (green curve).
        \label{S5-figures}}
\end{figure*}

We adopted for the position of S5 (\object{2MASS 14392160-6049528}, \object{WISE J143921.46-604952.9}) at epoch 2000.0 the coordinates from the 2MASS Point Source Catalog and not the positions we measure on the 2MASS Atlas image (Table~\ref{S-astrometry}).
The catalog coordinates are more accurate, since they are derived from the raw images \citepads{2006AJ....131.1163S}.
We note however that the difference between the position we measure in the image and the catalog value is very small, within $0.10\arcsec$.
The positional uncertainty listed in the 2MASS Point source catalog is  $\pm 0.08\arcsec$.
We derived a proper motion for S5 of $\mu_\alpha(\mathrm{S5}) = \PMRASfive \pm \PMerrSfive$\ mas\,yr$^{-1}$ and $\mu_\delta(\mathrm{S5}) = \PMdecSfive \pm \PMerrSfive$\ mas\,yr$^{-1}$.

We obtained a minimum approach distance of $\rho_\mathrm{min}(\mathrm{S5}) = \minSfive \pm \errminSfive\arcsec$ from $\alpha$\,Cen~A on {\dateSfive} $\pm 14$\,days (Fig.~\ref{S5-figures}, left panel).
Fortunately, the date of the closest approach of S5 is well within the observability period of $\alpha$\,Cen (January-August), and the two objects will therefore be in a favorable position in the sky.
In addition, the separation on sky of $\alpha$\,Cen A and B at the time of the event will be larger than $10\arcsec$, and the light contamination from $\alpha$\,Cen~B in the field of S5 will therefore be limited.
The magnitude difference in the $K$ band between $\alpha$\,Cen~A ($m_K = -1.5$) and S5 ($m_K = 7.8$) is $\Delta m_K = 9.3$ corresponding to a contrast of $f(\mathrm{S5})/f(\mathrm{A}) = 2\ 10^{-4}$.
This is the most favorable contrast for the stellar conjunctions with $\alpha$\,Cen within at least the next 30 years.

We obtained a satisfactory fit of the photometry of S5 considering an intrinsically red star with $T_\mathrm{eff}(\mathrm{S5}) = 3500 \pm 500$\,K, and a color excess of $E(B-V) = 6.5 \pm 1.0$ (Fig.~\ref{S5-figures}, right panel).
We excluded the WISE measurements in the W1 ($3.35\,\mu$m) as it is clear from the original WISE image that this measurement (reported by \citeads{2015AJ....149...64G}) is affected by contamination from $\alpha$\,Cen.
The WISE W4 ($22.1\,\mu$m) and MIPS $24\,\mu$m points show an excess compared to the adjusted SED, but it is uncertain if this excess is also due to contamination from $\alpha$\,Cen, so we did not take these points into account in the fit.
The very high value we obtained for $E(B-V)$ is consistent with the strong reddening expected for a distance of several kpc and the position in the galactic plane.
As a consequence of this particularly red color ($V-K = 13.7$), the predicted angular diameter is large with $\theta_\mathrm{LD}(\mathrm{S5}) = 0.47 \pm 0.05$\,mas.
These properties may indicate that S5 is a very distant red supergiant, for instance a twin of Betelgeuse ($\theta_\mathrm{LD} \approx 45$\,mas, $d\approx 200$\,pc) located at a distance of the order of 20\,kpc.
It could in principle also be an intrinsically smaller red giant star located at a few kiloparsec.
A photometric variability of this source is not excluded, and we foresee a thorough characterization in the coming years.

\subsubsection{Approach of $\alpha$\,Cen B to Star S6 in May 2031}

The minimum distance of $\alpha$\,Cen~B from S6 is $\rho_\mathrm{min}(\mathrm{S6}) = \minSsix \pm \errminSsix\arcsec$ on \dateSsix $\pm 28$\,days.
As for S4, the SED fit gives only limited information on the object due to the absence of infrared photometry.
We obtained an angular diameter $\theta_\mathrm{LD}(\mathrm{S6}) = 0.010 \pm 0.003$\,mas and an effective temperature of $T_\mathrm{eff}(\mathrm{S3}) = 4500 \pm 2000$\,K for $E(B-V)=0.5 \pm 0.5$.

\section{Potential of future stellar conjunctions\label{potential}}

\subsection{Relativistic deflection and amplification\label{deflection}}

Following \citetads{2014ApJ...782...89S} (see also \citeads{2008ApJ...684...59D}, \citeads{1996ARA&A..34..419P}), the angular radius $\theta_E$ of the Einstein ring of a point source is given by the expression:
\begin{equation}
\theta_E = \sqrt{\frac{4\,G M}{c^2 d_\pi}} \mathrm{\ \ with}\ 1/d_\pi= 1/D_L - 1/D_S.
\end{equation}
$M$ is the mass of the lens ($\alpha$\,Cen A or B in our case), $d_\pi$ the parallax distance of the lens, $D_L$ and $D_S$ the distances of the lens and of the source, respectively.
We can compute the Einstein ring radius of $\alpha$\,Cen A accurately as we know the mass of the star $M_A = 1.1055 \pm 0.0022\,M_\odot$ and its parallax $\pi_L = 747.17 \pm 0.61$\,mas from Sect~\ref{orbit}.
The parallaxes of the S stars cannot be estimated directly, but considering their high reddening, they are very likely negligible compared to that of $\alpha$\,Cen (S4 being possibly an exception).
Assuming that the approximation of a point source is valid for $\alpha$\,Cen~A
($\theta_\mathrm{LD} = 8.511 \pm 0.020$\,mas, from \citeads{2003A&A...404.1087K}),
we obtained an Einstein radius of $\theta_E(A) = 81.97 \pm 0.09$\,mas,
corresponding to approximately 20 times the angular radius of the photosphere of $\alpha$\,Cen~A (Fig.~\ref{S5-figures}).
The predicted impact parameter of the conjunction with S5 is $\rho_\mathrm{min} = \minSfive \pm \errminSfive\arcsec$
giving a probability of 45\% that S5 will enter the Einstein ring of $\alpha$\,Cen~A.
For $\alpha$\,Cen~B, $\theta_E(B)$ is estimated to $75.48 \pm 0.08$\,mas.

The gravitational lensing of the light from the background source results in the creation of two images, whose apparent shifts are given by \citepads{2014ApJ...782...89S}:
\begin{equation}
\theta_{\pm} = 0.5\,\left[u \pm \sqrt{u^2+4} \right]\,\theta_E,
\end{equation}
with $\theta_E$ the Einstein ring angular radius and $u=\rho_\mathrm{min} / \theta_E$.
Following the definition adopted in the rest of this article, $\rho_\mathrm{min}$ is the undeflected impact parameter (i.e.,~the minimum distance between the source and the lens in absence of gravitational lensing).
The photometric amplification factor $A$ is given by the expression:
\begin{equation}
A_\pm = 0.5\,\left[ \frac{u^2+2}{u \sqrt{4+u^2}} \pm 1\right].
\end{equation}
When the source is located at a large separation compared to $\theta_E$ ($u \gg 1$), the secondary image contribution is negligible. We thus consider only $A_+$ for S2, S3 and S4, for which the apparent shift $\delta \theta_+$ is between 2 and 5\,mas.

The deflection angles $\delta \theta_\pm$ and amplification factors $A_\pm$ for the conjunctions of $\alpha$\,Cen with stars S1 to S6 are listed in Table~\ref{conjunction-relativity}.
For these approximate computations, we consider only the gravitational lensing by the approached component of $\alpha$\,Cen and we neglect the influence of the other.
Due to the uncertainty on the coordinates and proper motions of S5 and S6, an almost perfect alignment with the lens cannot be excluded, and significantly higher amplification factors are possible.
For S5, the amplification could reach a significant factor for the sum of the two images (Table~\ref{conjunction-relativity}).
As we will precise the impact parameter in the future, we could in principle observe a near-perfect alignment and a very large gravitational amplification by positioning a spacecraft precisely where the impact parameter will be zero.
Shifting the apparent position of $\alpha$\,Cen~A on the sky by $0.05\arcsec$ requires to send an instrument at a distance of 0.067\,AU~$\approx 10$ million kilometers from the Earth (26 times the Earth-Moon distance).
The future availability of more accurate coordinates for S5 and the surrounding field (thanks e.g.,~to the \emph{Gaia} mission) will help in refining the value of the impact parameter at the Earth and the zero impact parameter projected trajectory in the solar system.

\begin{table*}[ht]
        \caption{Maximum apparent displacement $\delta \theta_\pm$ and amplification $A_\pm$ of the two images of S stars formed by the gravitational lens
        during the conjunctions with $\alpha$\,Cen A and B. The displacements are relative to the non-deflected (pre-lensing) positions.
        We also list the angular sizes $\theta_\mathrm{LD}$ of the S stars, and their projected linear diameters $\phi$ at $\alpha$\,Centauri.
	The error bars are given in super/subscripts position next to each value.
        }
        \centering          
        \label{conjunction-relativity}
        \begin{tabular}{ccccccrrrr}
	\hline\hline
        \noalign{\smallskip}
        Date & $L$$^a$ & $S$$^b$ & $\theta_\mathrm{LD}$[mas] & $\phi$[km] & $\rho_\mathrm{min}[\arcsec]$ & $\delta \theta_+[\mathrm{mas}]$ & $A_+[\%]$ & $\delta \theta_-[\mathrm{mas}]$ & $A_-[\%]$ \\
         \hline         
        \noalign{\smallskip}
         \dateSone & B & S1& $0.012_{0.002}$ & $2400_{400}$ & $\minSone \pm \errminSone$ & $+24.8^{+38.0}_{-10.4}$ & $101.2^{+92}_{-1.1}$ & $-230^{+139}_{-167}$ & $1.2^{+92}_{-1.1}$ \\
        \noalign{\smallskip}
         \dateStwo & A & S2 & $0.025_{0.003}$ & $5000_{600}$ & $\minStwo \pm \errminStwo$ & $+4.1^{+0.3}_{-0.3}$ & $100$ & - & - \\
        \noalign{\smallskip}
         \dateSthree & B & S3 & $0.018_{0.002}$ & $3600_{400}$ & $\minSthree \pm \errminSthree$ & $+4.2^{+0.7}_{-0.5}$ & $100$ & - & - \\
        \noalign{\smallskip}
         \dateSfour & B & S4 & $0.008_{0.003}$ & $1600_{600}$ & $\minSfour \pm \errminSfour$ & $+2.3^{+0.1}_{-0.1}$ & $100$ & - & - \\
        \noalign{\smallskip}
         \dateSfive & A & S5 & $0.47_{0.05}$ & $95000_{10000}$ & $\minSfive \pm \errminSfive$ & $+75^{+7}_{-39}$ & $321^{+\infty}_{-218}$ & $-90^{+8}_{-97}$  &  $221^{+\infty}_{-218}$ \\
        \noalign{\smallskip}
         \dateSsix & B & S6 & $0.010_{0.003}$ & $2000_{600}$ & $\minSsix \pm \errminSsix$ & $+20^{+56}_{-10}$ & $100.5^{+\infty}_{-0.4}$ & $-289^{+214}_{-267}$ & $0.5^{+\infty}_{-0.4}$ \\
        \noalign{\smallskip}
	\hline
        \end{tabular}
        \tablefoot{$^a$ $L$ is the lens (one of the $\alpha$\,Cen components).
        $^b$ $S$ is the lensed source (S stars).}
\end{table*}

\subsection{The close environment of $\alpha$\,Cen A and B}

The question of the stability of planetary orbits in the $\alpha$\,Cen system has been studied by several authors. The conclusion reached by \citetads{2016AJ....151..111Q} is that long-term stable orbits exist up to $3\arcsec$ (4\,AU) of each component, including the habitable zone \citepads{2014MNRAS.444.2167A}.
Stable circumbinary orbits are also found at distances larger than about $50\arcsec$ (70\,AU) from the center of mass, in agreement with \citetads{1997AJ....113.1445W}.
So exoplanets may exist around $\alpha$\,Cen, but their probability of presence is higher either close to each component or at large distances from the pair.
Through the photometric and astrometric monitoring of the S stars, we plan to probe the environment of $\alpha$\,Cen A and B for the presence of high and low mass planets, asteroids, comets or dust.
The $K$ band is particularly interesting for this type of observations as adaptive optics systems are very efficient in this wavelength range, providing excellent correction of the atmospheric turbulence.
Using a combination of coronagraphy and advanced processing techniques, the VLT/SPHERE instrument has demonstrated that sources with a contrast of $10^{-4}$ or smaller can be detected down to an angular separation of $\approx 3 \lambda/D = 0.2\arcsec$ with a high signal-to-noise ratio \citepads{2014A&A...572A..32P}.
It has also been proposed recently that optimized correction techniques may significantly improve the accessible contrast \citepads{2016160601895N}, particularly in view of the E-ELT instrumentation.
In addition to the photometric and astrometric observations, it will be very valuable to monitor spectroscopically the S stars and the $\alpha$\,Cen pair continuously before, during and after the conjunctions.

\subsubsection{Transits of dark low mass objects}

By analogy with stellar occultations by asteroids (e.g.~\citeads{2014Natur.508...72B}), the transit of very low mass objects (e.g.,~planets, comets or asteroids)
across gravitationally lensed images of the S sources will cause significant variations of their apparent flux.
The photometric stability of the approached stars will have to be checked before the event, but transits are expected to be achromatic and brief, facilitating their identification.
The determined angular diameters of the S1 to S6 sources range between 0.008\,mas (S4) and 0.47\,mas (S5), corresponding to projected linear diameters of
1600 to 95\,000\,km at $\alpha$\,Cen (Table~\ref{conjunction-relativity}).
Due to their small projected diameters, the probability to observe the transit of a planet in front of one of the S stars is low.
However, the conjunctions will allow us to efficiently search for the presence of the potentially more numerous asteroids and comets in the system through continuous monitoring of the photometry of the S stars.
Their small angular diameters decreases the probability of transit, but will result in a large relative change of the apparent brightness of the S star during the transit, even for a very small body.
For instance, the transit of a Ceres-sized asteroid ($\phi \approx 950$\,km) would induce a relative photometric change between 4\% on S2 and 30\% on S4 (only 0.01\% on S5).
Due to the extreme brightness of the $\alpha$\,Cen stars, the photometric monitoring of the conjunctions will be difficult for the fainter S stars at their closest approach.
Depending on the available instrumentation at the time of the transits, it may only be possible to follow their photometry down to a few arcseconds from the main stars.
They will in any case be useful to probe the circumbinary orbits domain, but thanks to the large parallax of $\alpha$\,Cen, even the inner regions within a few AU will be accessible.

The case of S5 is remarkable as its large linear diameter of 95\,000\,km projected at $\alpha$\,Cen~A results in a much more extended covered area in the $\alpha$\,Cen system during the approach.
The transit of an Earth-like planet ($\phi \approx 12\,700$\,km) would cause a drop of 1.8\% of the flux of S5.
As S5 is bright at infrared wavelengths, it will be possible to obtain highly accurate photometry down to very small angular separations from $\alpha$\,Cen~A.
In addition, the small impact parameter of the conjunction will result in the appearance of two amplified images while S5 evolves inside the Einstein ring of $\alpha$\,Cen~A.
These two images will move rapidly in the inner 0.1\,AU around the star and therefore probe a relatively large area within this short range.
They will be located symmetrically from $\alpha$\,Cen~A at angular separations around 80\,mas, within reach of the ALMA, E-ELT and VLTI/GRAVITY instruments.
When more accurate astrometry will be available for S5, we plan to conduct a detailed simulation of the conjunction taking into account the gravitational lensing from both $\alpha$\,Cen A and B.

\subsubsection{Secondary gravitational lensing by a massive planet}

As discussed by \citetads{2013ApJ...771...79D} and \citetads{2016ApJ...823..120N}, and more specifically in the case of Proxima by \citetads{2014ApJ...782...89S}, secondary gravitational lensing could occur due to the presence of a massive planet in the $\alpha$\,Cen system.
Such events are routinely observed in gravitational microlensing surveys (see e.g. \citeads{2015ApJ...804...33S}; \citeads{2016ApJ...822...75H}; \citeads{2006Natur.439..437B}).
As the mass of a planet is much lower than the mass of a star, the radius of its Einstein ring is also smaller.
But its dependence with the square root of the mass results in a non-negligible extension:
for a Jupiter-mass planet orbiting $\alpha$\,Cen, the angular radius of the Einstein ring is $\theta_E = 8$\,mas.
The VLTI/GRAVITY instrument will be able to measure the relative positions of $\alpha$\,Cen and the S stars
with an accuracy of $10\,\mu$as (\citeads{2011Msngr.143...16E}; \citeads{2014A&A...567A..75L}).
For a jovian mass planet, a deflection angle of $30\,\mu$as corresponds to an impact parameter of more than $2\arcsec$.
This means that the astrometric monitoring of the conjunctions with GRAVITY, independently of the impact parameter with the main
$\alpha$\,Cen components, gives us a $3\sigma$ sensitivity to the presence of a Jupiter-like planet over a $\pm 2\arcsec$ band
around the S star trajectory relatively to $\alpha$\,Cen.
Alternatively, the presence of a Neptune mass planet ($0.05\,M_J$) will be detectable within $\pm 0.1\arcsec$ from the relative
trajectory of the approached S star.
One potential difficulty of this observations is that the duration of the approach between the S star and the planet will be relatively short.
The proper motion of $\alpha$\,Cen is $3.7\arcsec\ \mathrm{yr}^{-1}$, so the crossing of the planet's Einstein ring will last less than two days.
But the deflection will be observable over a much longer period, including if the S star does not enter in the Einstein ring of the planet.
Considering a jovian mass planet, the duration during which the lensing event will be detectable will typically be of half a year.

\subsection{Proper motion of $\alpha$\,Cen, orbital parameters, parallax}

Due to its extreme apparent brightness, $\alpha$\,Cen's distance will not be measured accurately by the \emph{Gaia} mission.
But from high precision astrometry of the S star conjunctions, for instance with the GRAVITY instrument, it will be possible to derive its parallax with an accuracy of the order of a few $10\,\mu$as.
As star S5 is likely an evolved massive star, there are good prospects that it is observable in the radio domain with e.g. ALMA or VLBI.

The GRAVITY instrument can measure the differential position of two stars separated by up to $5\arcsec$ with the 1.8\,m VLTI Auxiliary Telescopes (ATs), and up to $2\arcsec$ with the 8\,m Unit Telescopes (UTs).
Thanks to its brightness, S5 will be observable with the ATs.
It will enter the $5\arcsec$ accessible range of GRAVITY in dual field approximately 18 months before the closest approach, and remain observable for another 18 months afterward (i.e.,~from late 2026 to late 2029).
It will be possible to follow the differential position of $\alpha$\,Cen~A and S5 with an accuracy of $10\,\mu$as per measurement epoch.
Over these three years, S5 will sweep through the area where stable planetary orbits were predicted by \citetads{2016AJ....151..111Q}.
Except for the approach of S4, the other conjunctions will be observable with the UTs in dual field mode.
Relativistic effects will have to be taken into account to correct the measured astrometric positions, but we already have accurate masses of $\alpha$\,Cen A and B with uncertainties of 0.4\% (Sect.~\ref{orbit}) so we can model the combined gravitational lens of the two stars with a high accuracy.
We also plan to obtain differential astrometry of $\alpha$\,Cen A and B in the coming years using GRAVITY to improve our estimate of the masses.

Following the conjunctions of $\alpha$\,Cen and the S stars will enable us to derive extremely accurate parallaxes for the components of $\alpha$\,Cen, potentially at the $10\,\mu$as level ($10^{-5}$ relative accuracy), that is, estimate their distances to $\pm 3$\,AU.
Combining the microarcsecond differential astrometry with ultra-precise radial velocities at a few cm\ s$^{-1}$ level obtained, for instance, using the frequency comb technology \citepads{2016NatCo...710436Y} will enable us to ascertain the orbital parameters and masses of $\alpha$\,Cen A and B with an exquisite accuracy.


\section{Conclusions}

We report that the two components of $\alpha$\,Cen will approach several distant stars within small angular separations in the coming decades.
An extraordinary conjunction of $\alpha$\,Cen A and the $m_K = 7.8$ star S5 will occur in early May 2028 at a minimum predicted separation of $\minSfive \pm \errminSfive\arcsec$.
These events provide remarkable opportunities to study the close environment of the two stars in ``transmission'' using the background approached star as a light probe.
During the approaches, astrometric measurements will reveal the relativistic deflection of the background star light with a high signal to noise ratio.
For the conjunction with S5, we may be able to directly observe the gravitational splitting of the distant source image using the E-ELT, VLTI/GRAVITY or ALMA.
The astrometric monitoring of the relative positions of $\alpha$\,Cen and the S stars may reveal the presence of planets through secondary gravitational lensing.
In addition, the light from the background star will possibly be subject to photometric variations induced by transits of low mass objects present in the $\alpha$\,Cen system.
Finally, following the S star conjunctions will give access to extremely accurate proper motion, orbital parameters and parallax values for $\alpha$\,Cen.
This accuracy will be valuable for high-precision modeling of the two stars of $\alpha$\,Cen, and for the preparation of the recently announced \emph{Breakthrough Starshot}\footnote{\url{https://breakthroughinitiatives.org/Initiative/3}} initiative to send ultra-fast light-driven nanocrafts to $\alpha$\,Centauri.

\begin{acknowledgements}
We thank ESO's Director General, Dr Tim de Zeeuw, for the allocation of Director's Discretionary Time with NACO to our program 296.D-5032, and the
Paranal staff for the successful execution of the observations.
This research received the support of PHASE, the partnership between ONERA, Observatoire de Paris, CNRS and University Denis Diderot Paris 7.
This research made use of Astropy\footnote{\url{http://www.astropy.org/}} \citepads{2013A&A...558A..33A}.
This research made use of the SIMBAD \citepads{2000A&AS..143....9W} and VIZIER databases (CDS, Strasbourg, France) and NASA's Astrophysics Data System.
This research has made use of the NASA/ IPAC Infrared Science Archive, which is operated by the Jet Propulsion Laboratory, California Institute of Technology, under contract with the National Aeronautics and Space Administration.
This paper makes use of the following ALMA data: ADS/JAO.ALMA\#2013.1.00170.S. ALMA is a partnership of ESO (representing its member states), NSF (USA), and NINS (Japan), together with NRC (Canada) and NSC and ASIAA (Taiwan), in cooperation with the Republic of Chile.
The Joint ALMA Observatory is operated by ESO, AUI/NRAO, and NAOJ.
This research has made use of the Washington Double Star Catalog maintained at the U.S. Naval Observatory.
\end{acknowledgements}

%
\bibliographystyle{aa} 
\bibliography{BiblioAlfCen.bib} 
%



%

\begin{appendix} 
\section{Relative astrometry of $\alpha$\,Cen A and B\label{relative-astrometry}}

The new and archival positions of $\alpha$\,Cen B relative to A are listed in Table~\ref{alfcen-relpos} since 1980.
The complete list of measurements is available from the Washington Double Star catalog
(\citeads{2016yCat....102026M}; \citeads{2001AJ....122.3466M}).
We adopted for each measurement the error bars indicated in the original paper, when they were available.
For the visual micrometer measurements, we adopted constant cartesian position uncertainties $\sigma = 0.10\arcsec$ between 1960 and 1976. 
We took $\sigma = 0.15\arcsec$ between 1930 and 1959, $0.20\arcsec$ between 1900 and 1929, $0.40\arcsec$ between 1850 and 1899, and $\sigma = 0.80\arcsec$ before 1850.
As a remark, the observation recorded in 1752 by Nicolas-Louis de Lacaille from Cape Town (South Africa) is well in agreement with the modern orbit, with deviations of only $d\alpha = 1.5 \pm 0.8\arcsec$ and $d\delta = 1.0 \pm 0.8\arcsec$ with respect to the expected position (see also \citeads{2014A&A...567A..26L}).

\begin{table}
        \caption{Relative position angle $\theta$ and separation $\rho$ of $\alpha$\,Cen A and B since 1980.}
        \centering          
        \label{alfcen-relpos}
        \begin{tabular}{crrl}
	\hline\hline
        \noalign{\smallskip}
        Epoch &  $\theta[^\circ]$ & $\rho[\arcsec]$ & Ref./\emph{Inst.}$^a$ \\
         \hline         
        \noalign{\smallskip}
2016.1893	& $	305.19	\pm	0.30	$ & $	4.013	\pm	0.02	$ & 	{\bf NACO}	\\
2015.3326	& $	293.30	\pm	0.60	$ & $	4.020	\pm	0.04	$ & 	{\bf ALMA B8}	\\
2014.9574	& $	288.28	\pm	0.70	$ & $	4.081	\pm	0.05	$ & 	{\bf ALMA B6}	\\
2014.5426	& $	282.91	\pm	0.15	$ & $	4.208	\pm	0.01	$ & 	{\bf ALMA B9}$^b$	\\
2014.5126	& $	282.84	\pm	0.30	$ & $	4.184	\pm	0.02	$ & 	{\bf ALMA B7}	\\
2014.2410	& $	279.20	\pm	0.30	$ & $	4.330	\pm	0.05	$ & 	An15	\\
2012.7100	& $	262.70	\pm	0.40	$ & $	5.050	\pm	0.05	$ & 	An14	\\
2009.7140	& $	244.20	\pm	0.40	$ & $	7.020	\pm	0.05	$ & 	An12	\\
2009.1517	& $	241.07	\pm	0.60	$ & $	7.400	\pm	0.07	$ & 	{\bf SOFI}	\\
2008.7190	& $	239.80	\pm	0.40	$ & $	7.780	\pm	0.05	$ & 	An11	\\
2007.3670	& $	235.70	\pm	0.60	$ & $	8.780	\pm	0.10	$ & 	An08	\\
2004.2739	& $	228.26	\pm	0.20	$ & $	11.122	\pm	0.02	$ & 	{\bf NACO}	\\
2002.6880	& $	224.30	\pm	1.00	$ & $	12.500	\pm	0.40	$ & 	An06	\\
2000.7400	& $	223.80	\pm	1.00	$ & $	13.300	\pm	0.40	$ & 	WDS	\\
2000.1950	& $	222.63	\pm	0.80	$ & $	13.933	\pm	0.20	$ & 	{\bf 2MASS}	\\
1991.2500	& $	215.29	\pm	0.15	$ & $	19.073	\pm	0.05	$ & 	Hip	\\
1989.5370	& $	214.68	\pm	0.16	$ & $	19.740	\pm	0.05	$ & 	J94	\\
1988.2880	& $	213.75	\pm	0.15	$ & $	20.180	\pm	0.05	$ & 	J94	\\
1987.2900	& $	213.21	\pm	0.15	$ & $	20.540	\pm	0.05	$ & 	J94	\\
1986.5090	& $	212.77	\pm	0.15	$ & $	20.784	\pm	0.05	$ & 	P91	\\
1985.6030	& $	212.08	\pm	0.15	$ & $	21.087	\pm	0.05	$ & 	P91	\\
1984.4250	& $	211.54	\pm	0.14	$ & $	21.385	\pm	0.05	$ & 	P91	\\
1982.4900	& $	210.60	\pm	0.40	$ & $	21.910	\pm	0.13	$ & 	T85	\\
1980.4000	& $	210.50	\pm	1.00	$ & $	21.710	\pm	0.50	$ & 	F86	\\
1980.2870	& $	210.04	\pm	0.06	$ & $	21.614	\pm	0.05	$ & 	J95	\\
	\hline
        \end{tabular}
        \tablefoot{$^a$ The instrument names in bold characters indicate
        new measurements reported in the present work.
        $^b$ The ALMA band 9 measurement is taken as the fiducial of our orbit determination (Sect.~\ref{orbit}). 
        The references are:
        An06: \citetads{2006JDSO....2..108A};
        An08: \citetads{2008JDSO....4...40A};
        An11: \citet{2011JDSO....7...64A};
        An12: \citetads{2012JDSO....8...15A};
	An14: \citetads{2014JDSO...10..232A};
	An15: \citetads{2015JDSO...11...81A};
        F86: \citet{Fr1986};
        Hip: \citetads{1997ESASP1200.....E};
        J94: \citetads{1994A&AS..107..235J};
        P91: \citetads{1991A&AS...88...63P};
        T85: \citetads{1985A&AS...59..449T};
        WDS: \citetads{2001AJ....122.3466M}.}
\end{table}

\section{Positions of stars S1 to S6}

We present in Tables~\ref{S-photometry} and \ref{S-astrometry} the photometric and astrometric measurements of the S stars S1 to S6, that we used to determine their angular diameters and place upper limits to their proper motions.

\begin{table}
        \caption{Photometry of stars S1 to S6.}
        \centering          
        \label{S-photometry}
        \begin{tabular}{llccl}
	\hline\hline
        \noalign{\smallskip}
        Band & $\lambda_0[\mu\mathrm{m}]$ & Mag. & Flux$^a$ & Ref.$^b$ \\
          \hline         
        \noalign{\smallskip}
        \multicolumn{5}{c}{\sc{Star S1}} \\
        Bessel V & 0.54 & $20.04 \pm 0.46$ & $3.5 \pm 1.6$  & K07  \\
        Bessel R & 0.64 & $19.35 \pm 0.34$ & $3.9 \pm 1.3$ & K07  \\
        Bessel I & 0.79 & $17.61 \pm 0.22$ & $10.2 \pm 2.2$ & K07  \\
        Bessel Z & 0.84 & $16.80 \pm 0.20$ & $21.0 \pm 4.2$ & K07  \\
        2MASS H & 1.62 & $14.13 \pm 0.31$ & $25.2 \pm 7.8$ & K07  \\
        2MASS K & 2.16 & $13.58 \pm 0.17$ & $15.8 \pm 2.7$ & K07  \\
        \hline         
        \noalign{\smallskip}
        \multicolumn{5}{c}{\sc{Star S2}} \\
        Bessel V & 0.54 & $15.66 \pm 0.20$ & $782 \pm 158$ & K07  \\
        Bessel R & 0.64 & $14.26 \pm 0.20$ & $1020 \pm 207$ & K07  \\
        Bessel I & 0.79 & $13.26 \pm 0.20$ & $973 \pm 197$ & K07  \\
        Bessel Z & 0.84 & $13.29 \pm 0.20$ & $965 \pm 195$ & K07  \\
        2MASS J & 1.25 & $11.97 \pm 0.04$ & $510 \pm 19$ & K07  \\
        2MASS H & 1.62 & $11.38 \pm 0.05$ & $320 \pm 15$ & K07  \\
        2MASS K & 2.16 & $11.14 \pm 0.03$ & $150 \pm 4.2$ & K07  \\
	IRAC 5.8 & 5.8 & $10.59 \pm 0.20$ &  $6.0 \pm 1.2$ & S09 \\
	IRAC 8.0 & 8.0 & $10.30 \pm 0.20$ & $2.3 \pm 0.4$ & S09 \\
        \hline         
        \noalign{\smallskip}
        \multicolumn{5}{c}{\sc{Star S3}} \\
        Bessel V & 0.54 & $19.72 \pm 0.35$ & $4.7 \pm 1.8$ & K07  \\
        Bessel R & 0.64 & $17.57 \pm 0.21$ & $20.3 \pm 4.3$ & K07  \\
        Bessel I & 0.79 & $16.04 \pm 0.20$ & $43.5 \pm 8.8$ & K07  \\
        Bessel Z & 0.84 & $15.74 \pm 0.20$ & $55.6 \pm 11$ & K07  \\
        2MASS J & 1.25 & $13.90 \pm 0.19$ & $86.2 \pm 17$ & K07  \\
        2MASS H & 1.62 & $12.85 \pm 0.08$ & $82.1 \pm 6.3$ & K07  \\
        2MASS K & 2.16 & $12.89 \pm 0.11$ & $29.9 \pm 3.2$ & K07  \\
        \hline         
        \noalign{\smallskip}
        \multicolumn{5}{c}{\sc{Star S4}} \\
        Bessel V & 0.54 & $17.51 \pm 0.20$ & $35.9 \pm 7.3$ & K07  \\
        Bessel R & 0.64 & $16.59 \pm 0.20$ & $50.1 \pm 10$ & K07  \\
        Bessel I & 0.79 & $15.95 \pm 0.20$ & $47.2 \pm 10$ & K07  \\
        Bessel Z & 0.84 & $16.22 \pm 0.20$ & $35.8 \pm 7.2$ & K07  \\
         \hline         
        \noalign{\smallskip}
        \multicolumn{5}{c}{\sc{Star S5}} \\
        Bessel V & 0.54 & $21.53 \pm 0.95$ & $0.9 \pm 0.9$ & K07  \\
        Bessel R & 0.64 & $20.00 \pm 0.41$ & $2.2 \pm 1.0$ & K07  \\
        Bessel I & 0.79 & $18.66 \pm 0.27$ & $3.9 \pm 1.1$ & K07  \\
        Bessel Z & 0.84 & $18.00 \pm 0.22$ & $6.9 \pm 1.6$ & K07  \\
        2MASS J & 1.25 & $11.60 \pm 0.03$ & $717 \pm 20$ & K07  \\
	2MASS H & 1.62 & $9.125 \pm 0.027$ & $2540 \pm 64$ & G15 \\
	2MASS K$_s$ & 2.16 & $7.756 \pm 0.024$ & $3380 \pm 76$ & G15 \\
	IRAC 3.6 & 3.6 & $7.161 \pm 0.058$  & $888 \pm 49$ & G15 \\
	IRAC 4.5 & 4.5 & $6.912 \pm 0.052$ & $458 \pm 22$ & G15 \\
	WISE W2 & 4.6 & $6.089 \pm 0.025$ & $886 \pm 21$ & G15 \\
	IRAC 5.8 & 5.8 & $6.500 \pm 0.027$ &  $258 \pm 6.4$ & G15 \\
	IRAC 8.0 & 8.0 & $6.257 \pm 0.027$ & $94.5 \pm 2.5$ & G15 \\
	MSX6C A & 8.28 & $6.05 \pm 0.06$ & $89.4 \pm 4.7$ & E03 \\
	WISE W3 & 11.6 & $5.725 \pm 0.017$ & $33.4 \pm 0.5$\ & G15 \\
	WISE W4 & 22.1 & $3.924 \pm 0.029$ & $13.7 \pm 0.4$ & G15 \\
	MIPS 24 & 24 & $4.22 \pm 0.02$ & $7.8 \pm 0.2$ & G15 \\
        \hline         
        \noalign{\smallskip}
        \multicolumn{5}{c}{\sc{Star S6}} \\
        Bessel V & 0.54 & $16.93 \pm 0.20$ & $41.2 \pm 12$ & K07  \\
        Bessel R & 0.64 & $15.98 \pm 0.20$ & $87.8 \pm 18$ & K07  \\
        Bessel I & 0.79 & $15.38 \pm 0.20$ & $79.8 \pm 16$ & K07  \\
        Bessel Z & 0.84 & $15.65 \pm 0.20$ & $60.4 \pm 12$ & K07  \\
        \hline
        \end{tabular}
\tablefoot{$^a$ The flux is expressed in units of $10^{-16}\ \mathrm{W\ m}^{-2}\ \mu$m$^{-1}$. $^b$ E03: \citetads{2003yCat.5114....0E}; G15: \citetads{2015AJ....149...64G}; K07: \citetads{2007A&A...464..373K}; S09: \citetads{2009yCat.2293....0S}.}
\end{table}

\begin{table*}
        \caption{Astrometry of the stars approached by $\alpha$\,Cen A and B until 2031.
        The listed coordinates are referenced to the 2MASS frame (ICRS).}
        \centering          
        \label{S-astrometry}
        \begin{tabular}{lcccccl}
	\hline\hline
        \noalign{\smallskip}
        Date & Band & MJD$^a$ & RA & Dec & $\sigma_\mathrm{rel} [\arcsec]^b$ & Instrument \\
         \hline         
        \noalign{\smallskip}
       \multicolumn{7}{c}{\sc{Star S1}} \\
2000-03-12 & $K_s$ &  51615.383 &  14:39:25.277 &  -60:49:49.57 & 0.30 & 2MASS$^c$\\
2004-02-29 & $I$  &  53064.271 &   14:39:25.228 &  -60:49:49.60 & 0.05 & SUSI2\\ 
2004-04-02 & $Z$ &  53097.121 &  14:39:25.224 &  -60:49:49.60 & 0.05 & SUSI2\\
2004-04-02 & $I$  &  53097.243  &  14:39:25.223 & -60:49:49.64 & 0.05 & SUSI2\\ 
2009-02-25 & $J_s$ &  54887.356 &  14:39:25.224 & -60:49:49.67 & 0.05 & SOFI \\
         \hline         
        \noalign{\smallskip}
       \multicolumn{7}{c}{\sc{Star S2}} \\
2000-03-12 & $K_s$ &  51615.383 &  14:39:24.291 &  -60:49:53.66 & 0.20 & 2MASS$^c$\\
2004-02-26 & $V$ &  53061.325 &  14:39:24.284 &  -60:49:53.79 & 0.10 & SUSI2\\
2004-02-29 & $I$  &  53064.271 &   14:39:24.285 &  -60:49:53.80 & 0.05 & SUSI2\\
2004-02-29 & $V$  &  53064.329 &  14:39:24.284 &  -60:49:53.77 & 0.10 & SUSI2\\
2004-04-02 & $Z$ &  53097.121 &  14:39:24.284 &  -60:49:53.83 & 0.05 & SUSI2\\
2004-04-02 & $V$ &  53097.181 &  14:39:24.287 &  -60:49:53.77 & 0.10 & SUSI2\\
2004-04-02 & $I$  &  53097.243  &  14:39:24.284 &  -60:49:53.75 & 0.05 & SUSI2\\
2009-02-25 & $J_s$ &  54887.356 &  14:39:24.288 &  -60:49:53.80 & 0.05 & SOFI \\
2016-03-27 & $K_s$ &  57474.405 &  14:39:24.287 &  -60:49:53.80 & 0.06 & NACO\\
         \hline         
        \noalign{\smallskip}
       \multicolumn{7}{c}{\sc{Star S3}} \\
2000-03-12 & $K_s$ &  51615.383 &  14:39:24.279 &  -60:49:46.03 & 0.20 & 2MASS$^c$\\
2004-02-25 & $R$ &  53060.277 &  14:39:24.271 &  -60:49:46.02 & 0.05 & SUSI2\\
2004-02-29 & $I$  &  53064.271 &   14:39:24.266 & -60:49:46.05 & 0.05 & SUSI2\\
2004-04-02 & $Z$ &  53097.121 &  14:39:24.264 &  -60:49:46.07 & 0.05 & SUSI2\\
2004-04-02 & $I$  &  53097.243  &  14:39:24.269 &  -60:49:46.07 & 0.05 & SUSI2\\
2009-02-25 & $J_s$ &  54887.356 &  14:39:24.269 & -60:49:46.07 & 0.05 & SOFI \\
         \hline         
        \noalign{\smallskip}
       \multicolumn{7}{c}{\sc{Star S4}} \\
2004-02-25 & $R$ &  53060.277 &  14:39:23.472 &  -60:49:42.97 & 0.05 & SUSI2\\
2004-02-29 & $I$  &  53064.271 &   14:39:23.482 & -60:49:42.97 & 0.05 & SUSI2\\
2004-02-29 & $V$  &  53064.329 &  14:39:23.477 &  -60:49:42.97 & 0.10 & SUSI2\\
2004-04-02 & $Z$ &  53097.121 &  14:39:23.467 &  -60:49:43.04 & 0.05 & SUSI2\\
2004-04-02 & $V$ &  53097.181 &  14:39:23.482 &  -60:49:42.97 & 0.10 & SUSI2\\
2004-04-02 & $I$  &  53097.243  &  14:39:23.477 &  -60:49:42.97 & 0.05 & SUSI2\\
2009-02-25 & $J_s$ &  54887.356 &  14:39:23.481 & -60:49:42.89 & 0.05 & SOFI \\
2016-03-27 &  $K_s$ &  57474.405 &  14:39:23.485 & -60:49:42.91 & 0.06 &  NACO\\
         \hline 
        \noalign{\smallskip}
       \multicolumn{7}{c}{\sc{Star S5}} \\
2000-03-12 &  $K_s$ &  51615.383 &  14:39:21.600 &  -60:49:52.84 & 0.08 & 2MASS$^d$\\
2004-04-02 &  $Z$ &  53097.121 &  14:39:21.589 &  -60:49:52.82 & 0.05 & SUSI2\\
2009-02-25 &  $J_s$ &  54887.356 &  14:39:21.585 &  -60:49:52.93 & 0.05 & SOFI\\
2016-03-27 &  $K_s$ &  57474.405 &  14:39:21.586 &  -60:49:52.89 & 0.06 & NACO\\
         \hline 
        \noalign{\smallskip}
       \multicolumn{7}{c}{\sc{Star S6}} \\
2004-02-25 & $R$ &  53060.277 &  14:39:20.523 &  -60:49:41.41 & 0.05 & SUSI2\\
2004-02-29 & $I$  &  53064.271 &   14:39:20.528 & -60:49:41.37 & 0.05 & SUSI2\\
2004-02-29 & $V$  &  53064.329 &  14:39:20.528 &  -60:49:41.37 & 0.10 & SUSI2\\
2004-04-02 &  $Z$ &  53097.121 &  14:39:20.513 &  -60:49:41.48 & 0.05 &  SUSI2\\
2004-04-02 & $V$ &  53097.181 &  14:39:20.523 &  -60:49:41.41 & 0.10 & SUSI2\\
2004-04-02 & $I$  &  53097.243  &  14:39:20.523 &  -60:49:41.37 & 0.05 & SUSI2\\
2009-02-25 &  $J_s$ &  54887.356 &  14:39:20.518 &  -60:49:41.41 & 0.05 &  SOFI\\
        \hline
        \end{tabular}
        \tablefoot{$^a$ MJD is the average modified julian date of the measurement.
        $^b$ $\sigma_\mathrm{rel}$ is the position uncertainty with respect to the 2MASS frame in arcseconds, including statistical uncertainty and systematic offset between instruments (see Sect.~\ref{referencing}).
        $^c$ The coordinates are measured on the 2MASS Atlas image in the $K_s$ band.
        $^d$ The coordinates of S5 are taken from the 2MASS Point Source catalog.
        }
\end{table*}

\end{appendix}
\end{document}